\documentclass[10pt,final,doublecolumn]{IEEEtran}
\usepackage{amsmath}
\usepackage{amssymb}
\usepackage{amsfonts}
\usepackage{latexsym}
\usepackage{graphicx}
\usepackage{bbding}
\usepackage{enumerate}
\usepackage{subeqnarray}
\usepackage{multicol}
\usepackage{color}
\usepackage{setspace}
\usepackage{mathrsfs}
\usepackage{array}
\usepackage{algorithm,algorithmic}
\usepackage{colortbl}
\usepackage{bm}
\IEEEoverridecommandlockouts
\allowdisplaybreaks[4]

\expandafter\def\expandafter\normalsize\expandafter{%
    \normalsize%
    \setlength\abovedisplayskip{2.5pt}%
    \setlength\belowdisplayskip{2.5pt}%
    \setlength\abovedisplayshortskip{1.2pt}%
    \setlength\belowdisplayshortskip{1.2pt}%
}

\begin{document}
\title{Continuous Aperture Array-Assisted Integrated Communication and Navigation in LEO Satellite Constellations}
\author{
Qi Wang, Xiaoming Chen, Qiao Qi, Zhaolin Wang, and Yuanwei Liu
\thanks{Qi Wang and Xiaoming Chen are with the College of Information Science and Electronic Engineering, Zhejiang University, Hangzhou 310027, China (e-mails: wang-qi@zju.edu.cn; chen\_xiaoming@zju.edu.cn). Qiao Qi is with the School of Information Science and Technology, Hangzhou Normal University, Hangzhou 311121, China (e-mail: qiqiao@hznu.edu.cn).
Zhaolin Wang and Yuanwei Liu are with the Department of Electrical and Electronic Engineering, the University of Hong Kong, Hong Kong, China (e-mail: zhaolin.wang@hku.hk; yuanwei@hku.hk).
}
}\maketitle

\begin{abstract}
This paper proposes a novel continuous aperture array (CAPA)-assisted integrated communication and navigation (ICAN) framework for low Earth orbit (LEO) satellite constellations. Within this framework, an electromagnetic-based collaborative transmission model is developed, in which multiple satellites equipped with CAPAs simultaneously radiate downlink data streams and navigation reference signals over shared spectrum. Building upon this, the achievable communication rate and the navigation Cramer-Rao bound (CRB) are derived, which explicitly characterize the intrinsic coupling between the dual-function beamformers and system performance. To improve the positioning accuracy with communication quality of service guarantee, a joint beamforming optimization problem is formulated to minimize the average CRB subject to transmit power budgets and minimum rate constraints. To tackle the inherent infinite-dimensionality of the CAPA beamformer design, an ICAN channel subspace is introduced to equivalently transform the formulation into a tractable finite-dimensional problem, which is then efficiently solved via an iterative convex optimization algorithm. Finally, numerical results demonstrate that the proposed CAPA-assisted beamforming design algorithm significantly outperforms conventional discrete phased array architectures and other benchmark schemes, yielding notable improvements in ICAN performance.

\end{abstract}


\section{Introduction}
The sixth-generation (6G) mobile communication era is driving the evolution toward globally connected, where low Earth orbit (LEO) satellite constellations, as non-terrestrial network (NTN), are expected to play an important role \cite{LEO satellites}. By deploying thousands of satellites worldwide, these mega-constellations aim to provide space-based connectivity with seamless coverage, low latency, and high throughput \cite{NTN}.

Beyond communication, dense LEO satellite constellations also offer an attractive platform for high-performance positioning, navigation, and timing (PNT) services. Compared with traditional global navigation satellite systems (GNSS) operating in medium and high orbits, LEO satellite constellations can provide improved geometry and signal conditions due to their orbital characteristics and enhanced visibility, which enables stronger geometric diversity and a more robust signal environment \cite{GNSS}. These features have motivated integrated communication and navigation (ICAN) as an emerging direction for 6G networks \cite{ICAN1}, \cite{ICAN2}. By jointly designing communication and navigation functions on a shared platform, ICAN can improve spectrum and hardware utilization, reduce deployment cost and payload complexity, and deliver performance gains for both services.

However, achieving high-performance ICAN is constrained by conventional antenna technologies, particularly spatially discrete phased arrays that are widely used in current satellite payloads. The high orbital velocity of LEO satellites requires agile and continuous beam tracking. In practice, discrete arrays suffer from scan-angle-dependent gain loss, quantization errors, and non-negligible calibration overhead, which makes it difficult to meet these requirements \cite{scan-angle-dependent}. To alleviate these limitations, holographic multiple-input multiple-output (MIMO) and dynamic metasurface architectures have been explored by using densely packed elements to approximate a continuous aperture \cite{holographic MIMO}, \cite{Dynamic Metasurface}. Nevertheless, these solutions still rely on discrete element control and typically cannot realize a truly continuous current distribution. As discussed in \cite{degrees of freedom}, this prevents them from fully exploiting the electromagnetic degrees of freedom offered by the aperture. Moreover, ICAN imposes concurrent beamforming requirements, including highly directional multiuser beams for communication and broad, structured radiation patterns for navigation, which further increases the implementation complexity and power consumption of discrete architectures.

To address these limitations, a continuous aperture array (CAPA) architecture is introduced as an appealing solution. CAPA is rooted in electromagnetic information theory, which studies the fundamental limits of wireless systems by directly modeling radiating surfaces and wave propagation \cite{electromagnetic-based}. Unlike discrete phased arrays, a CAPA is modeled as a continuous radiating surface that supports a spatially continuous current distribution over the aperture, thereby providing additional degrees of freedom for electromagnetic field synthesis. The authors in \cite{CAPA spatial degrees} demonstrated that CAPAs can achieve high spatial degrees of freedom, and the authors in \cite{CAPA channel capacity} investigated the corresponding channel capacity, showing notable gains over discrete arrays of comparable size. Building on these foundations, the authors in \cite{CAPA multi-user} developed a CAPA beamforming framework for multiuser downlink transmission, further illustrating the design flexibility enabled by continuous apertures. Collectively, these research findings suggest that CAPA is a promising enabler for ICAN, as it can support high-gain directional transmission for communication and can potentially accommodate the wide-coverage, structured radiation required for navigation.

Despite this potential, several challenges remain in designing CAPA-assisted ICAN systems. While recent efforts have started to explore beamforming for CAPA-enabled dual-functional transmission, most existing studies consider a single CAPA radiator deployed on a single platform. For instance, in \cite{CAPA ISAC1}, the authors investigated downlink and uplink integrated sensing and communication (ISAC) transmission in CAPA systems and developed corresponding beamforming designs for dual-functional operation. In \cite{CAPA ISAC2}, the authors studied CAPA-enabled ISAC from a rate-Cramer-Rao bound (CRB) perspective and proposed a beamforming framework to characterize and optimize the inherent sensing-communication trade-off. In addition to this system-level limitation, existing works also face a methodological challenge in solving continuous-aperture beamforming problems. A common approach is to discretize the continuous-aperture beamforming problem via Fourier-series expansions. For example, the authors in \cite{CAPA ref1} approximated the continuous source current distribution using a finite set of Fourier basis functions, which turns the original functional optimization into a high-dimensional finite-dimensional problem, and the required basis size can grow rapidly with the aperture size and carrier frequency. To avoid such discretization-induced accuracy-complexity issues, the authors in \cite{CAPA ref2} developed a calculus-of-variations-based design, where closed-form updates are derived within an iterative weighted minimum mean-squared error (WMMSE) framework to directly optimize the continuous beamformers without relying on discretization approximations. However, for NTN scenarios such as LEO satellite constellations, these single-platform formulations do not readily extend to cooperative multi-satellite transmission with multiple CAPAs, where satellite propagation channels differ fundamentally from terrestrial channels. Consequently, how to systematically exploit multiple CAPAs in a LEO satellite constellation for ICAN, while balancing communication throughput and navigation accuracy, remains largely unexplored.

Motivated by the aforementioned research gaps, this paper develops a CAPA-assisted ICAN framework for LEO satellite constellations and proposes a tractable subspace-based approach for joint beamforming design. The main contributions are summarized as follows:

\begin{enumerate}
\item We propose a CAPA-assisted ICAN architecture for LEO satellite constellations with cooperative multi-satellite transmission, and establish an electromagnetic-based continuous-aperture model to derive the achievable rates of communication user equipments (CUEs) and the CRBs of navigation user equipments (NUEs).

\item We formulate a joint beamforming design problem that minimizes the average CRB of NUEs subject to per-satellite transmit power budgets and minimum achievable rate constraints for all CUEs, thereby characterizing the fundamental accuracy-throughput trade-off in ICAN.

\item We introduce an ICAN channel subspace to transform the infinite-dimensional CAPA beamforming design into an equivalent finite-dimensional formulation, based on which we develop an efficient iterative convex optimization algorithm to obtain feasible solutions.
\end{enumerate}

The remainder of this paper is organized as follows. Section II presents the system model and performance metrics. Section III formulates the joint beamforming optimization problem and develops the proposed subspace-based solution algorithm. Section IV provides numerical results, and Section V concludes the paper.

\emph{Notations}: Scalars, vectors, and matrices are denoted by regular letters, boldface lowercase letters, and boldface uppercase letters, respectively. For complex-valued quantities, $(\cdot)^{*}$, $(\cdot)^{T}$, and $(\cdot)^{H}$ represent the complex conjugate, transpose, and conjugate transpose, respectively, and $\Re\{\cdot\}$ extracts the real part. The matrix inverse is denoted by $(\cdot)^{-1}$. The Euclidean norm of a vector is written as $\|\cdot\|$, and $\times$ denotes the cross product of two three-dimensional vectors. For a matrix, $\mathrm{tr}(\cdot)$ and $\mathrm{rank}(\cdot)$ denote the trace and rank, respectively. The block-diagonal operator is denoted by $\mathrm{blkdiag}(\cdot)$, and $\mathbf{X}\succeq \mathbf{0}$ indicates that $\mathbf{X}$ is positive semidefinite. The sets of complex- and real-valued $a\times b$ matrices are denoted by $\mathbb{C}^{a\times b}$ and $\mathbb{R}^{a\times b}$, respectively. The asymptotic computational complexity is expressed using the notation $\mathcal{O}(\cdot)$.

\section{System Model}
\begin{figure}[!t] \centering
\includegraphics [width=0.49\textwidth] {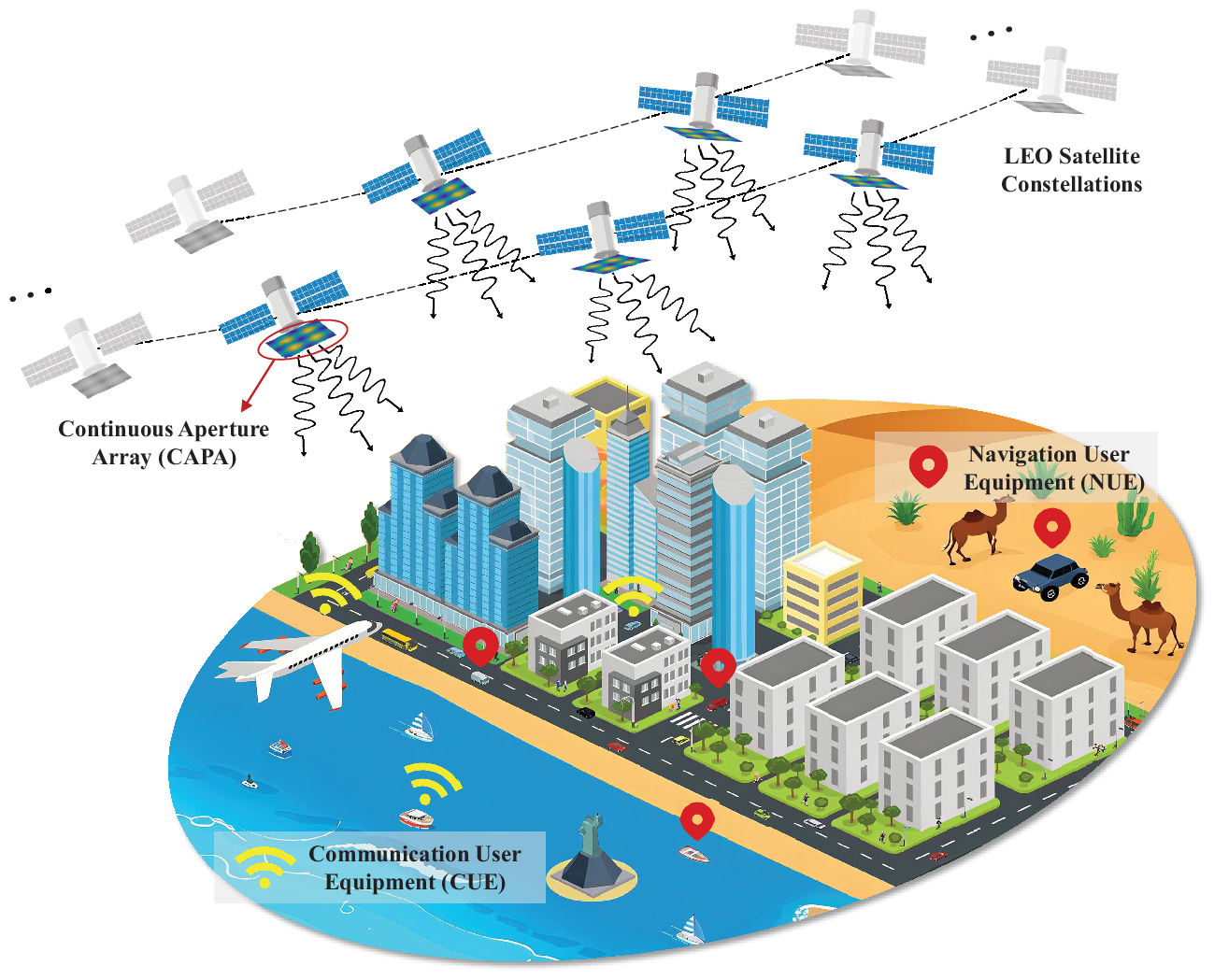}
\caption {System model for CAPA-assisted ICAN in LEO satellite constellation.}
\label{system model}
\end{figure}
Consider a LEO satellite constellation-based ICAN system, as shown in Fig. \ref{system model}. The constellation follows a Walker Delta configuration, comprising a large number of LEO satellites arrayed in multiple equally inclined circular orbital planes with evenly spaced right ascension of the ascending node (RAAN) and uniform in-plane phasing. Each satellite is equipped with a geocentric nadir-pointing CAPA transmitter that radiates electromagnetic waves supporting both data communication and position navigation over a shared spectrum, realized via a continuous and precisely controllable sinusoidal source current distribution across its radiating surface. Without loss of generality, at any given time, a subset of $K$ satellites is dynamically selected as a collaborative service group to jointly serve $M$ single-antenna CUEs and $L$ single-antenna NUEs within their terrestrial coverage area. Enabled by high-speed, low-latency inter-satellite links (ISLs), the satellites in the service group exchange channel state information (CSI) and ephemeris data in real time, enabling coordinated multi-satellite joint transmission. Through the satellite-terrestrial links, CUEs decode modulated data streams for global connectivity, whereas NUEs exploit dedicated navigation reference signals to achieve high-precision positioning. 

To enable a rigorous mathematical description, we first establish the coordinate frames. Here, the Earth-centered, Earth-fixed (ECEF) frame is adopted as the global Cartesian reference, with its origin at the Earth's center of mass, the $x$-axis pointing toward the intersection of the equator and the Greenwich meridian, the $y$-axis pointing eastward along the equator, and the $z$-axis aligned with the rotation axis toward the North Pole. In this context, the positions of the $k$-th satellite, the $m$-th CUE, and the $l$-th NUE can be denoted by vectors $\mathbf{r}_k = [r_k^x, r_k^y, r_k^z]^T$, $\mathbf{p}_m = [p_m^x, p_m^y, p_m^z]^T$, and $\mathbf{q}_l = [q_l^x, q_l^y, q_l^z]^T$, respectively, all expressed in the ECEF frame. In parallel, for each satellite, a local orbital frame (LOF) is defined with its origin at the CAPA surface center, $x'$-axes equal to $y' \times z'$ forming a right-handed triad, $y'$-axes along the instantaneous velocity, and $z'$-axes pointing toward Earth's center. The transformation from the LOF to the global ECEF frame can be realized by a rotation matrix $\mathbf{R}_k \in \mathbb{R}^{3 \times 3}$, which is uniquely determined by the satellite's ECEF position $\mathbf{r}_k$ and velocity $\mathbf{v}_k$.
In this context, the CAPA on the $k$-th satellite is modeled as a rectangular surface $\mathcal{S}'_k$ lying in the $x'$-$y'$ plane of its LOF, i.e.,
\begin{equation}
\mathcal{S}'_k \!=\! \left\{ \mathbf{s}'_k \!=\! (s_k^{'x}, s_k^{'y}, 0)^T \;\middle|\; |s_k^{'x}| \le \frac{D_x}{2},\; |s_k^{'y}| \le \frac{D_y}{2} \right\},
\end{equation}
where $D_x$ and $D_y$ are the aperture side lengths, and $|\mathcal{S}'_k|= D_x D_y = A_T$, with $A_T$ denoting the effective radiating area of the CAPA. Herein, $\mathbf{s}'_k \in \mathcal{S}'_k$ denotes the LOF coordinates of an arbitrary point on the CAPA of the $k$-th satellite.
Accordingly, the ECEF position $\mathbf{s}_k$ of any point on this surface is obtained by mapping its LOF coordinate $\mathbf{s}'_k$ with the satellite's pose, and is given by
\begin{equation}
\mathbf{s}_k = \mathbf{r}_k + \mathbf{R}_k \mathbf{s}'_k.
\label{eq:global_pos}
\end{equation}
This geometric formulation provides a precise description of the spatial location and attitude of the $k$-th satellite's CAPA, forming the basis for the transmit, propagation, communication and navigation models.

\subsection{Transmit Model}
To simultaneously serve multiple CUEs for downlink communications and multiple NUEs for terrestrial navigation, the $k$-th satellite transmits a superposition of $M$ downlink data streams and one dedicated navigation reference signal. Let $x_m^{\mathrm{C}}$ denote the complex data symbol intended for the $m$-th CUE, which is modeled as an independent, zero-mean circularly symmetric complex Gaussian random variable with normalized power, i.e., $\mathbb{E}\!\left[|x_m^{\mathrm{C}}|^2\right]=1$.
Likewise, let $x_k^{\mathrm{N}}$ represent the navigation reference signal emitted by the $k$-th satellite, modeled as a deterministic unit-power pseudo-random sequence, i.e., $\langle |x_k^{\mathrm{N}}|^2\rangle_T=1$, and time-balanced such that $\langle x_k^{\mathrm{N}}\rangle_T=0$, where $\langle \cdot \rangle_T \triangleq \frac{1}{T}\int_0^T (\cdot)\,dt$ denotes time averaging over a sufficiently long interval $T$. The known pseudo-random waveform $x_k^{\mathrm N}$ serves as a receiver-side spreading code reference to extract position and Doppler-dependent features via matched filtering.
Moreover, the random data symbols $\{x_m^{\mathrm{C}}\}$ and the deterministic navigation signal $\{x_k^{\mathrm{N}}\}$ are assumed to be mutually uncorrelated. To radiate these scalar signals into physical space via a CAPA, a mapping from the signal domain to the continuous aperture domain is required.

Unlike conventional discrete arrays that apply complex weights to a finite set of antenna elements, a CAPA enables continuous spatial modulation of the surface current across the aperture.
Accordingly, we characterize the CAPA excitation corresponding to the above superposed baseband signals by a surface current density distribution over the continuous aperture. Specifically, let $\tilde{\mathbf{J}}(\mathbf{s}_k,\varpi)\in\mathbb{C}^{3\times 1}$ denote the Fourier transform of the source current density at an ECEF coordinate $\mathbf{s}_k$, where $\varpi=2\pi f/c$ is the spatial angular frequency, with $f$ being the carrier frequency and $c$ being the speed of light. Since all satellites operate on the same carrier $f$, we omit the explicit dependence on $\varpi$ and write $\tilde{\mathbf{J}}(\mathbf{s}_k)$ for brevity. Because the CAPA is mounted on the satellite platform, the current distribution is most naturally specified in the LOF. We focus on a uni-polarized CAPA, where the current is excited exclusively along the LOF $y'$-axis, represented by $\hat{\mathbf{u}}'_y=[0,1,0]^T$. The corresponding polarization direction in the ECEF frame is $\hat{\mathbf{u}}_{k,y}=\mathbf{R}_k\hat{\mathbf{u}}'_y$. Under this setting, the source current vector in ECEF can be expressed as
\begin{equation}
\tilde{\mathbf{J}}(\mathbf{s}_k)= j(\mathbf{s}'_k)\,\hat{\mathbf{u}}_{k,y},
\end{equation}
where $j(\mathbf{s}'_k)$ (in ampere per meter, [A/m]) is the scalar surface current field that captures the complex amplitude-phase distribution over the continuous CAPA aperture $\mathcal{S}'_k$ in LOF coordinates.

Leveraging the linearity of electromagnetic radiation with respect to the source, we model the scalar current field as a linear superposition of communication and navigation-bearing components, which is given by
\begin{equation}\label{source current}
j(\mathbf{s}'_k)=\sum_{m=1}^{M} w_{k,m}(\mathbf{s}'_k)\,x_m^{\mathrm{C}} + v_k(\mathbf{s}'_k)\,x_k^{\mathrm{N}}.
\end{equation}
Herein, $w_{k,m}(\mathbf{s}'_k)$ is the continuous transmit beamformer representing the source current pattern at $\mathbf{s}'_k$ tailored for the $m$-th CUE, and $v_{k}(\mathbf{s}'_k)$ is the continuous transmit beamformer designed for navigation services.

Based on these properties, the total average transmit power at the $k$-th satellite is derived by taking both the statistical and time averages of the instantaneous power, which is expressed as
\begin{align}
&P_k= \mathbb{E} \left[ \left\langle \int_{\mathcal{S}'_k} \left| j(\mathbf{s}'_k) \right|^2 d\mathbf{s}'_k \right\rangle_T \right] \nonumber \\
&\overset{(\mathrm{a})}= \int_{\mathcal{S}'_k} \left( \sum_{m=1}^{M} |w_{k,m}(\mathbf{s}'_k)|^2 \mathbb{E}[|x_m^C|^2] + |v_{k}(\mathbf{s}'_k)|^2 \langle|x_k^N|^2\rangle_T \right) d\mathbf{s}'_k \notag\\
&\overset{(\mathrm{b})}= \sum_{m=1}^{M} \int_{\mathcal{S}'_k} |w_{k,m}(\mathbf{s}'_k)|^2 d\mathbf{s}'_k + \int_{\mathcal{S}'_k} |v_{k}(\mathbf{s}'_k)|^2 d\mathbf{s}'_k. \label{eq:power_derivation_b}
\end{align}
The equality in (a) follows from the uncorrelated nature of all distinct signal components, which makes the cross terms average to zero. The equality in (b) follows directly from the unit power normalization of both communication and navigation signals.

\subsection{Propagation Model}
The electromagnetic field radiated by the source current $\mathbf{J}(\mathbf{s}_k)$ on the CAPA surface propagates through the atmosphere in a homogeneous medium to the ground users. According to Maxwell's equations, the total electric field at the location $\mathbf{p}_m$ of the $m$-th CUE, due to the $K$ cooperating satellites, is given by the superposition of surface integrals over each satellite's aperture, i.e., \cite{Large Intelligent Surfaces}
\begin{equation}
\mathbf{E}_m^C = \sum_{k=1}^K \int_{\mathcal{S}'_k} \mathbf{G}_{k,m}^C\left( \mathbf{s}_k(\mathbf{s}'_k), \mathbf{p}_m \right) \tilde{\mathbf{J}}(\mathbf{s}_k(\mathbf{s}'_k)) d\mathbf{s}'_k.
\end{equation}
Similarly, the electric field experienced by the $l$-th NUE at location $\mathbf{q}_l$ is expressed as
\begin{equation}
\mathbf{E}_l^N = \sum_{k=1}^K \int_{\mathcal{S}'_k} \mathbf{G}_{k,l}^N\left( \mathbf{s}_k(\mathbf{s}'_k), \mathbf{q}_l \right) \tilde{\mathbf{J}}(\mathbf{s}_k(\mathbf{s}'_k)) d\mathbf{s}'_k.
\end{equation}
Note that the integration is carried out over the local coordinate $\mathbf{s}'_k$ on the surface $\mathcal{S}'_k$, whereas the integrand is evaluated at the corresponding global ECEF position $\mathbf{s}_k(\mathbf{s}'_k)$.
Furthermore, the integral kernel $\mathbf{G}_{k,m}^C(\cdot)$ and $\mathbf{G}_{k,l}^N(\cdot)$ are dyadic Green's functions for line-of-sight (LoS) propagation. In the far-field region, where the electromagnetic field has attained its radiating form, $\mathbf{G}_{k,m}^C(\cdot)$ is well-approximated by \cite{integral kernel}
\begin{align}
\mathbf{G}_{k,m}^{C}\left( \mathbf{s}_k, \mathbf{p}_m \right) = - &
\frac{j\eta e^{ -j\frac{2\pi}{\lambda}\left\| \mathbf{p}_m - \mathbf{s}_k \right\|}}{2\lambda \left\| \mathbf{p}_m - \mathbf{s}_k \right\|}
\xi_r^{\frac{1}{2}} e^{-j\psi_{k,m}} e^{-j\phi^{\mathrm{D}}_{k,m}} \notag\\
&\cdot \left( \mathbf{I}_3 - \frac{(\mathbf{p}_m - \mathbf{s}_k)(\mathbf{p}_m - \mathbf{s}_k)^T}{\left\| \mathbf{p}_m - \mathbf{s}_k \right\|^2} \right),
\label{eq:green_function}
\end{align}
where $\eta = 120\pi  \Omega$ is the intrinsic impedance of free space, $\lambda$ denotes the signal wavelength, and $\mathbf{I}_3$ is the $3 \times 3$ identity matrix. Moreover, rain attenuation is captured by the amplitude-phase pair $(\xi_r^{1/2}, \psi_{k,m})$, with the amplitude factor $\xi_r$ modeled as log-normal so that $\ln(\xi_r^{1/2}) \sim \mathcal{N}(\mu_r,\sigma_r^2)$ and $\psi_{k,m}$ denoting the associated rain-induced phase perturbation \cite{rain attenuation}.
The term $e^{-j\phi^{\mathrm{D}}_{k,m}}$ denotes the Doppler shift caused by the LEO satellite motion, which can be estimated and compensated by standard receiver synchronization such as carrier tracking and is omitted in the subsequent analysis \cite{Doppler}. The Green's function for the navigation link $\mathbf{G}_{k,l}^N(\mathbf{s}_k, \mathbf{q}_l)$ has an identical mathematical structure.

In order to sense the full three-dimensional electric field, each CUE and NUE would require an ideal tri-polarized receiver, which is challenging to implement in practice. Accordingly, we assume that each user employs a uni-polarized antenna. For the $m$-th CUE, the receive polarization is described by a real unit vector $\mathbf{u}_m \in \mathbb{R}^{3 \times 1}$ with $\|\mathbf{u}_m\| = 1$, so that only the field component along $\mathbf{u}_m$ is observed. Without loss of generality, all CUE antennas are assumed to be $y$-directed linearly polarized in ECEF, i.e., $\mathbf{u}_m = [0, 1, 0]^T$. To facilitate electromagnetic propagation analysis, we define an effective spatially continuous channel response from an arbitrary point $\mathbf{s}'_{k}$ on the $k$-th satellite's CAPA to the $m$-th CUE. This channel response accounts for the dyadic Green's function as well as transmit and receive polarization effects, and is given by
\begin{equation}
h_{k,m}^{C}\left( \mathbf{s}'_k, \mathbf{p}_m \right) =  \mathbf{u}_m^T \mathbf{G}_{k,m}^C\left( \mathbf{s}_k(\mathbf{s}'_k), \mathbf{p}_m \right) \hat{\mathbf{u}}_{k,y}.
\label{eq:comm_channel_response}
\end{equation}
Similarly, the spatially continuous channel response from the same radiating point to the $l$-th NUE is given by
\begin{equation}
h_{k,l}^{N}\left( \mathbf{s}'_k, \mathbf{q}_l \right) =  \mathbf{u}_l^T \mathbf{G}_{k,l}^N\left( \mathbf{s}_k(\mathbf{s}'_k), \mathbf{q}_l \right) \hat{\mathbf{u}}_{k,y},
\label{eq:nav_channel_response}
\end{equation}
where $\mathbf{u}_l \in \mathbb{R}^{3 \times 1}$ denotes the unit receive polarization vector of the $l$-th NUE, defined in the same manner as $\mathbf{u}_m$.
These channel response functions provide a rigorous and concise foundation for formulating the system performance metrics for both communication and navigation, as detailed in the following subsections.

\subsection{Communication Model}
With the effective channel responses established, we now formulate the signal reception model for the communication service. The total signal received by the $m$-th CUE is the coherent superposition of the electromagnetic waves radiated by the service group of $K$ satellites, corrupted by additive thermal noise. By invoking the source current definition in equation (\ref{source current}) and the continuous channel response $h_{k,m}^{\mathrm{C}}(\mathbf{s}'_k, \mathbf{p}_m)$ derived in equation (\ref{eq:comm_channel_response}), the received signal at the $m$-th CUE $y_{m}^{\mathrm{C}}$ can be expressed as
\begin{align}
&y_m^C
= \sum_{k=1}^K \int_{\mathcal{S}'_k} h_{k,m}^C(\mathbf{s}'_k, \mathbf{p}_m)\, j(\mathbf{s}'_k)\, d\mathbf{s}'_k + n_m^C \nonumber \\
&= \underbrace{ \left( \sum_{k=1}^K \int_{\mathcal{S}'_k} h_{k,m}^C(\mathbf{s}'_k, \mathbf{p}_m)\, w_{k,m}(\mathbf{s}'_k)\, d\mathbf{s}'_k \right) x_m^C }_{\text{Desired communication signal}}
\nonumber\\
&\ \  \ + \underbrace{ \sum_{i\ne m}^M \left( \sum_{k=1}^K \int_{\mathcal{S}'_k} h_{k,m}^C(\mathbf{s}'_k, \mathbf{p}_m)\, w_{k,i}(\mathbf{s}'_k)\, d\mathbf{s}'_k \right) x_i^C }_{\text{Inter-CUE interference}}
\nonumber\\
&\ \  + \underbrace{ \sum_{k=1}^K \left( \int_{\mathcal{S}'_k} h_{k,m}^C(\mathbf{s}'_k, \mathbf{p}_m)\, v_k(\mathbf{s}'_k)\, d\mathbf{s}'_k \right) x_k^N }_{\text{Navigation interference}}
\;+\; \underbrace{ n_m^C }_{\text{Noise}}.
\label{eq:comm_received_signal_decomposed}
\end{align}
The first term is the desired communication signal, namely the coherent superposition of transmissions from all $K$ cooperating satellites intended for the $m$-th CUE. The integral $\sum_{k=1}^K \int_{\mathcal{S}'_k} h_{k,m}^C(\cdot) w_{k,m}(\cdot) d\mathbf{s}'_k$ acts as the effective cooperative channel gain from the service group to $\mathbf{p}_m$. The next two terms constitute interference. Inter-CUE interference arises from energy radiated by the continuous transmit beamformers $\{w_{k,i}(\mathbf{s}'_k)\}$ designed for other CUEs $(i\neq m)$ that leaks toward the $m$-th CUE. Navigation interference is generated by the navigation continuous transmit beamformer $\{v_k(\mathbf{s}'_k)\}$, an intrinsic coupling of the ICAN architecture. The last term $n_m^C$ denotes thermal noise at the $m$-th CUE, modeled as independent additive white Gaussian noise (AWGN) with zero mean and variance $\sigma_{C,m}^2$.

Based on the decomposition in equation \eqref{eq:comm_received_signal_decomposed} and the statistical independence of the transmitted symbols, the signal-to-interference-plus-noise ratio (SINR) received at the $m$-th CUE is given by
\begin{equation}
\gamma_m = \frac{ \left| \sum_{k=1}^K \int_{\mathcal{S}'_k} h_{k,m}^C(\mathbf{s}'_k, \mathbf{p}_m) w_{k,m}(\mathbf{s}'_k) d\mathbf{s}'_k \right|^2 }{ I_{m,\text{CUE}} + I_{m,\text{NAV}} + \sigma_{C,m}^2 },
\label{eq:sinr}
\end{equation}
where the interference powers are defined as
\begin{align}
I_{m,\text{CUE}} &= \sum_{i \ne m}^M \left| \sum_{k=1}^K \int_{\mathcal{S}'_k} h_{k,m}^C(\mathbf{s}'_k, \mathbf{p}_m) w_{k,i}(\mathbf{s}'_k) d\mathbf{s}'_k \right|^2, \\
I_{m,\text{NAV}} &= \sum_{k=1}^K \left| \int_{\mathcal{S}'_k} h_{k,m}^C(\mathbf{s}'_k, \mathbf{p}_m) v_k(\mathbf{s}'_k) d\mathbf{s}'_k \right|^2.
\end{align}
Finally, the achievable rate of the $m$-th CUE is determined by the Shannon-Hartley theorem, given by
\begin{equation}
R_m = \log_2(1 + \gamma_m),
\label{eq:rate}
\end{equation}
where $R_m$ is measured in bits per second per Hertz (bps/Hz). This formulation explicitly captures the coupling effect between communication and navigation signals and characterizes how the transmit beamformers $w_{k,m}(\mathbf{s}_k)$ and $v_k(\mathbf{s}_k)$ jointly determine the achievable rate.

\subsection{Navigation Model}
Consider the $l$-th NUE located at position $\mathbf{q}_l$. The received signal comprises the desired navigation component, superimposed communication transmissions, and receiver noise.
To obtain a satellite-specific navigation observation, the NUE applies a matched filter using the known pseudo-random reference waveform $x_k^{\rm N}$, which suppresses navigation signals from other satellites. After this satellite-specific matched filtering, by substituting the channel response defined in equation \eqref{eq:nav_channel_response}, the resulting observation at the $l$-th NUE associated with the $k$-th satellite can be written as
\begin{align} \label{receive_navigation}
y_{k,l}^{\rm N}
&=
\underbrace{
\int_{\mathcal S_k'}
h^{\rm N}_{k,l}(\mathbf s_k',\mathbf q_l)\,
v_k(\mathbf s_k')\, {\rm d}\mathbf s_k'
}_{\text{Desired navigation signal}}
+
n_{k,l}^{\rm eff},
\end{align}
where $n_{k,l}^{\rm eff}$ represents the effective interference-plus-noise term at the $l$-th NUE for the $k$-th satellite after matched filtering, which is given by
\begin{align}
n_{k,l}^{\rm eff}
&=
\underbrace{
\sum_{k'=1}^{K}
\sum_{m=1}^{M}
\int_{\mathcal S_{k'}'}
h^{\rm N}_{k',l}(\mathbf s_{k'}',\mathbf q_l)\,
w_{k',m}(\mathbf s_{k'}')\,
{\rm d}\mathbf s_{k'}' \rho'_{k,m}
}_{\text{Communication interference}}
+
n_{k,l}^N,
\label{eq:neff_def}
\end{align}
where $\rho'_{k,m}= \langle x_m^{\rm C}\,x_k^{{\rm N}*}\rangle_T$ is the
communication residual coefficient after the matched filter, and $n_{k,l}^N$ denotes the AWGN at the $l$-th NUE with zero mean and variance $\sigma_{N,l}^2$.

We adopt a direct position estimation (DPE) framework, where the NUE's position $\mathbf{q}_l$ is estimated directly from the received signals \cite{GNSS_DPE}. To fully leverage the geometric diversity provided by the satellite constellation, we model the aggregate received signal as a $K$-dimensional vector, formed by stacking the distinct observations associated with each satellite, i.e.,
\begin{align}
\mathbf y_l^{\rm N}
&=
\big[y_{1,l}^{\rm N},\,y_{2,l}^{\rm N},\,\ldots,\,y_{K,l}^{\rm N}\big]^{T}
\in \mathbb C^{K\times 1},
\label{eq:vec_nav_obs}
\end{align}
which can be compactly expressed as
\begin{align}
\mathbf y_l^{\rm N}
=
\boldsymbol\mu_l(\mathbf q_l)
+
\mathbf n_l^{\rm eff},
\label{eq:vec_nav_model}
\end{align}
where $\mathbf y_l^{\rm N}$ is the observed navigation measurement vector, $\boldsymbol\mu_l(\mathbf q_l)\in\mathbb C^{K\times 1}$ is the position-dependent noise-free mean vector collecting the desired navigation components, and $\mathbf n_l^{\rm eff}\in\mathbb C^{K\times 1}$ is the effective interference-plus-noise vector. The $k$-th entry of $\boldsymbol\mu_l(\mathbf q_l)$ in equation \eqref{eq:vec_nav_model} is given
by
\begin{align}
\mu_{k,l}(\mathbf q_l)
&=
\int_{\mathcal S_k'}
h^{\rm N}_{k,l}(\mathbf s_k',\mathbf q_l)\,
v_k(\mathbf s_k')\, {\rm d}\mathbf s_k',
\label{eq:mu_kl_def}
\end{align}
and we have
$\boldsymbol\mu_l(\mathbf q_l) = [\mu_{1,l}(\mathbf q_l),\ldots,\mu_{K,l}(\mathbf q_l)]^{T}$.
Due to the superposition of a large number of independent communication signals after the matched filter, the central limit theorem motivates modeling the aggregate interference as a complex Gaussian process. Consequently, the effective interference-plus-noise is modeled as
\begin{align}
\mathbf n_l^{\rm eff}
\sim \mathcal{CN}\big(\mathbf 0,\sigma_{{\rm eff},l}^2\,\mathbf I_K\big),
\label{eq:noise_cov}
\end{align}
where $\mathbf I_K$ is the $K\times K$ identity matrix, and the effective noise variance $\sigma_{{\rm eff},l}^2$ is given by
\begin{equation}\label{effective_noise_variance}
\sigma_{\mathrm{eff},l}^2 = \eta^\rho \sum_{m=1}^M \left| \sum_{k=1}^K \int_{\mathcal{S}'_k} h_{k,l}^N(\mathbf{s}'_k, \mathbf{q}_l) w_{k,m}(\mathbf{s}'_k) d\mathbf{s}'_k \right|^2 + \sigma_{N,l}^2,
\end{equation}
where $\eta^\rho = \mathbb E[|\rho'_{k,m}|^2]$ with the assumption that $\eta^\rho$ is identical for all $k$ and $m$.
Based on this model, the probability density function (PDF) of the observation $\mathbf y_l^{\rm N}$ given the position $\mathbf{q}_l$ is expressed as \cite{GNSS_MLE}
\begin{equation}
p(\mathbf y_l^{\rm N}|\mathbf{q}_l) = \frac{1}{(\pi \sigma_{{\rm eff},l}^2)^K}
\exp\!\left(
-\frac{\big\|\mathbf y_l^{\rm N} - \boldsymbol\mu_l(\mathbf q_l)\big\|^2}
{\sigma_{{\rm eff},l}^2}
\right).
\label{y_PDF}
\end{equation}
Derived from equation (\ref{y_PDF}), the constant-free negative log-likelihood, which serves as the cost function, is written as
\begin{equation}
\mathcal L(\mathbf q_l)
=
\frac{\big\|\mathbf y_l^{\rm N} - \boldsymbol\mu_l(\mathbf q_l)\big\|^2}
{\sigma_{{\rm eff},l}^2}
+
K\log\sigma_{{\rm eff},l}^2.
\label{eq:nll_exact_final}
\end{equation}
In practice, the dependence of $\sigma_{{\rm eff},l}^2$ on the position $\mathbf q_l$ and the beamformers is relatively weak compared to that of the vector $\boldsymbol\mu_l(\mathbf q_l)$. Following standard practice in deterministic parameter estimation, we adopt an iterative scheme in which $\sigma_{{\rm eff},l}^2$ is updated using the beamformers and the previous position estimate, and then treated as a constant within each iteration. Under this approximation,
minimizing $\mathcal L(\mathbf q_l)$ is equivalent to minimizing the squared error term $\big\|\mathbf y_l^{\rm N} - \boldsymbol\mu_l(\mathbf q_l)\big\|^2$.
To enable gradient-based position refinement, we define the Jacobian matrix of the vector $\boldsymbol\mu_l$ with respect to the position $\mathbf q_l$ as
\begin{equation}
\mathbf J_l(\mathbf q_l)=\frac{\partial \boldsymbol\mu_l(\mathbf q_l)}{\partial \mathbf q_l^{T}}=[\nabla_{\mathbf q_l}\mu_{1,l}(\mathbf q_l),\ldots,\nabla_{\mathbf q_l}\mu_{K,l}(\mathbf q_l)]^{T},
\label{eq:J_def}
\end{equation}
where the gradient of the $k$-th component is given by
\begin{align}
\nabla_{\mathbf q_l} \mu_{k,l}(\mathbf q_l)
&=
\int_{\mathcal S_k'}
\nabla_{\mathbf q_l}
h^{\rm N}_{k,l}(\mathbf s_k',\mathbf q_l)\,
v_k(\mathbf s_k')\, {\rm d}\mathbf s_k',
\label{eq:grad_mu_kl}
\end{align}
and the closed-form expression of $\nabla_{\mathbf q_l} h^{\rm N}_{k,l}(\mathbf s_k',\mathbf q_l)$ is provided in Appendix~A. By using equation \eqref{eq:J_def}, the gradient of the squared error term is obtained as
\begin{equation}
\nabla_{\mathbf q_l}
\big\|\mathbf y_l^{\rm N} - \boldsymbol\mu_l(\mathbf q_l)\big\|^2
=
-2\,\Re\Big\{
\mathbf J_l^{H}(\mathbf q_l)
\big(\mathbf y_l^{\rm N} - \boldsymbol\mu_l(\mathbf q_l)\big)
\Big\}.
\label{eq:grad_cost}
\end{equation}
Given an initial position estimate
$\widehat{\mathbf q}_l^{(0)}$, a gradient-descent based update can be applied as
\begin{align}
\widehat{\mathbf q}_l^{(t+1)}
&=
\widehat{\mathbf q}_l^{(t)}
-
\alpha^{(t)}
\nabla_{\mathbf q_l}
\big\|\mathbf y_l^{\rm N} - \boldsymbol\mu_l(\mathbf q_l)\big\|^2
\Big|_{\mathbf q_l = \widehat{\mathbf q}_l^{(t)}},
\label{eq:GD_update}
\end{align}
where $\alpha^{(t)} > 0$ denotes the step size at the $t$-th iteration. Once convergence, the $l$-th NUE can obtain its position $\widehat{\mathbf q}_l$.

To quantify the fundamental limit of positioning accuracy, we employ the CRB as the navigation performance metric, which provides a lower bound on the variance of any unbiased estimator. Under the complex Gaussian observation model, the Fisher information matrix (FIM) with respect to the position vector $\mathbf{q}_l$ is given by \cite{FIM}
\begin{equation} \label{eq:FIM_def}
\mathbf{F}_l =
\frac{2}{\sigma_{{\rm eff},l}^2}\,
\mathrm{Re}\Big\{
\mathbf J_l^{H}(\mathbf q_l)\,
\mathbf J_l(\mathbf q_l)
\Big\}.
\end{equation}
The CRB matrix is defined as the inverse of the FIM, i.e.,
\begin{align}
\mathbf{CRB}(\mathbf q_l)
&=
\mathbf F_l^{-1}(\mathbf q_l).
\label{eq:CRB_def}
\end{align}
Since the diagonal elements of the CRB matrix correspond to the variance bounds of the individual coordinates, its trace $\mathrm{tr}(\mathbf{CRB}(\mathbf q_l))$ serves as a scalar metric quantifying the lower bound on the total mean-squared position error (MSE) for the $l$-th NUE.
Crucially, this expression reveals that the ultimate navigation accuracy is intrinsically coupled with the beamforming design for both communication $w_{k,m}(\mathbf{s}_k)$ and navigation $v_k(\mathbf{s}_k)$, as these functions determine both the signal gradient and the effective noise variance.

\section{Subspace-Based CAPA-ICAN Beamforming Design in LEO Satellite Constellations}
In this section, building on the preceding derivations and analysis, we show that system performance is fully characterized by the spatially continuous source current distributions over the CAPA surfaces of all $K$ serving satellites. In particular, the transmit beamformers for communication $w_{k,m}(\mathbf{s}_k)$ and for navigation $v_k(\mathbf{s}_k)$ jointly determine the quality delivered to CUEs and NUEs. Therefore, we present a beamforming design for ICAN to improve the overall performance in LEO satellite constellations.

\subsection{Problem Formulation}
To enhance the overall performance of a dual-function LEO satellite constellation, we minimize the average trace of the CRB for position estimation while simultaneously guaranteeing a minimum achievable rate for all CUEs and satisfying the per-satellite maximum transmit power constraints on each CAPA surface by optimizing the transmit beamformers. The optimization problem is mathematically stated as follows:
\begin{subequations}
\label{op1}
\begin{align}
&\min_{ \{w_{k,m}\}, \{v_k\} } \quad \frac{1}{L} \sum_{l=1}^L \mathrm{tr}\left( \mathbf{CRB}(\mathbf{q}_l) \right) \label{op1obj} \\
&\mathrm{s.t.} \quad \log_2(1 + \gamma_m) \ge R_m^{\min}, \label{op1st1} \\
& \quad \int_{\mathcal{S}'_k} \! \left( \sum_{m=1}^M |w_{k,m}(\mathbf{s}'_k)|^2 + |v_k(\mathbf{s}'_k)|^2 \right) \! d\mathbf{s}'_k \le P_k^{\max}, \label{op1st2}
\end{align}
\end{subequations}
In this formulation, the objective function \eqref{op1obj} seeks to minimize the average positioning error bound across all NUEs. The constraint \eqref{op1st1} based on equation (\ref{eq:rate}) ensures that each CUE attains its required minimum QoS threshold with $R_m^{\min}$ being the required minimum achievable rate. The constraint \eqref{op1st2} derived from equation (\ref{eq:power_derivation_b}) enforces a per-satellite transmit power limit at the CAPA surface with $P_k^{\max}$ being the current factor proportional to the maximum transmit power budget.
It is worth noting that the primary challenge in solving \eqref{op1} stems from its infinite-dimensional nature, since the optimization variables are functions defined over the continuous domain of the CAPA surfaces. To render this problem tractable, it is necessary to transform it into a finite-dimensional equivalent without sacrificing optimality. Moreover, the quadratic terms in the optimization variables and the cross-coupling induced by the dual-function design render problem \eqref{op1} non-convex and generally intractable, so an exact optimal solution cannot be obtained in polynomial time. In this context, we develop an efficient algorithm that obtains feasible suboptimal solutions and improves the overall performance of the CAPA-assisted ICAN system in LEO satellite constellations. The subsequent subsection focuses on a dimensionality-reduction transformation of the problem and the corresponding algorithm design.

\subsection{Channel Subspace Transformation}
Because the optimization variables are defined over the continuous surfaces $\mathcal{S}'_k$, we adopt a channel subspace-based approach projecting the infinite-dimensional problem onto a finite-dimensional subspace without loss of optimality \cite{subspace}. The key to dimensionality reduction is to restrict the search space of the transmit beamformers to a properly constructed joint communication and navigation channel subspace.

\emph{Theorem 1:} Define the finite-dimensional ICAN channel subspace for the $k$-th satellite, spanned by the basis functions given by the conjugates of the continuous communication and navigation channel responses associated with the CAPA surface as
\begin{equation}\label{ICAN_channel_subspace}
\mathcal{C}_k \triangleq \mathrm{span} \left( \{h_{k,m}^C(\mathbf{s}'_k, \mathbf{p}_m)^*\}_{m=1}^M \cup \{h_{k,l}^N(\mathbf{s}'_k, \mathbf{q}_l)^*\}_{l=1}^L \right),
\end{equation}
where $\text{span}(\cdot)$ denotes the set of all functions formed by finite linear combinations of the listed basis functions.
Then there exists an optimal solution to problem \eqref{op1} whose transmit beamformers satisfy $w_{k,m}(\mathbf{s}'_k) \in \mathcal{C}_k $ for all $m$ and $v_k(\mathbf{s}'_k)\in \mathcal{C}_k $.

\begin{IEEEproof}
Please see Appendix B.
\end{IEEEproof}

According to Theorem 1, an optimal set of transmit beamformers $w_{k,m}(\mathbf{s}'_k)$ and $v_k(\mathbf{s}'_k)$ can be chosen to lie entirely within $\mathcal{C}_k $, denoted as
\begin{align} \label{subspace_w}
w_{k,m}(\mathbf{s}'_k)
&= \sum_{m'=1}^{M} \alpha_{k,m,m'}^{C}\, h_{k,m'}^{C}(\mathbf{s}'_k,\mathbf{p}_{m'})^{*} \nonumber\\
&\quad + \sum_{l'=1}^{L} \beta_{k,m,l'}^{C}\, h_{k,l'}^{N}(\mathbf{s}'_k,\mathbf{q}_{l'})^{*} \nonumber\\
&= \boldsymbol{\Phi}_k^{T}(\mathbf{s}'_k)\, \mathbf{a}_{k,m}.
\end{align}
\begin{align} \label{subspace_v}
v_k(\mathbf{s}'_k)
&= \sum_{m=1}^{M} \alpha_{k,m}^{N}\, h_{k,m}^{C}(\mathbf{s}'_k,\mathbf{p}_{m})^{*}
+ \sum_{l=1}^{L} \beta_{k,l}^{N}\, h_{k,l}^{N}(\mathbf{s}'_k,\mathbf{q}_{l})^{*} \notag\\
&= \boldsymbol{\Phi}_k^{T}(\mathbf{s}'_k)\, \mathbf{b}_{k},
\end{align}
where $ \mathbf{a}_{k,m}
= \Big[\, \alpha_{k,m,1}^{C},\ldots,\alpha_{k,m,M}^{C},\, \beta_{k,m,1}^{C},\ldots,\beta_{k,m,L}^{C} \,\Big]^{T}
\in \mathbb{C}^{(M+L)\times 1}$, $\mathbf{b}_{k}
= \Big[\, \alpha_{k,1}^{N},\ldots,\alpha_{k,M}^{N},\, \beta_{k,1}^{N},\ldots,\beta_{k,L}^{N} \,\Big]^{T}
\in \mathbb{C}^{(M+L)\times 1}$, and the subspace basis vector $\boldsymbol{\Phi}_k(\mathbf{s}'_k)$ for the $k$-th satellite is denoted as
\begin{align}
&\boldsymbol{\Phi}_k(\mathbf{s}'_k)
=\Big[\, h_{k,1}^{C}(\mathbf{s}'_k,\mathbf{p}_1)^{*},\ldots,h_{k,M}^{C}(\mathbf{s}'_k,\mathbf{p}_M)^{*},\,\notag\\
&\ \ \ \ h_{k,1}^{N}(\mathbf{s}'_k,\mathbf{q}_1)^{*},\ldots,h_{k,L}^{N}(\mathbf{s}'_k,\mathbf{q}_L)^{*} \,\Big]^{T}\in\mathbb{C}^{(M+L)\times 1}.
\end{align}
Herein, $\alpha_{k,m,m'}^{C}$ and $\beta_{k,m,l'}^{C}$ are the expansion coefficients of the communication transmit beamformer $w_{k,m}$ on the conjugate communication and navigation channel bases, respectively, and $\alpha_{k,m}^{N}$ and $\beta_{k,l}^{N}$ are the corresponding coefficients of the navigation transmit beamformer $v_k$.

By leveraging the subspace transformation, the infinite-dimensional optimization variables $w_{k,m}(\mathbf{s}'_k)$ and $v_k(\mathbf{s}'_k)$ are parameterized by the finite-dimensional coefficient vectors $\mathbf{a}_{k,m}$ and $\mathbf{b}_k$, respectively.
In this context, by substituting equations \eqref{subspace_w} and \eqref{subspace_v} into equations \eqref{eq:comm_received_signal_decomposed} and \eqref{receive_navigation}, the received signals at the $m$-th CUE and the $l$-th NUE can be rewritten as
\begin{equation}
y_m^C = \sum_{k=1}^K \Big( \mathbf{g}_{k,m}^{C}\mathbf{a}_{k,m} x_m^C
+ \sum_{i\neq m} \mathbf{g}_{k,m}^{C}\mathbf{a}_{k,i} x_i^C
+ \mathbf{g}_{k,m}^{C}\mathbf{b}_k x_k^N \Big) + n_m^C,
\end{equation}
\begin{equation}
y_{k,l}^N =  \mathbf{g}_{k,l}^{N} \mathbf{b}_k + \sum_{k'=1}^{K} \sum_{m=1}^M \mathbf{g}_{k',l}^{N} \mathbf{a}_{k',m} \rho'_{k,m} + n_{k,l}^N,
\end{equation}
where $\mathbf{g}_{k,m}^{C}\in \mathbb{C}^{1\times (M+L)}$ and $\mathbf{g}_{k,l}^{N} \in \mathbb{C}^{1\times (M+L)}$ denote the effective subspace channel vectors from the $k$-th satellite to the $m$-th CUE and to the $l$-th NUE, respectively, defined as
\begin{equation}
\mathbf{g}_{k,m}^{C} = \int_{\mathcal{S}'_k} h_{k,m}^{C}(\mathbf{s}'_k,\mathbf{p}_m)\,\boldsymbol{\Phi}_k^{T}(\mathbf{s}'_k)\, d\mathbf{s}'_k,
\end{equation}
\begin{equation}
\mathbf{g}_{k,l}^{N} = \int_{\mathcal{S}'_k} h_{k,l}^{N}(\mathbf{s}'_k,\mathbf{q}_l)\,\boldsymbol{\Phi}_k^{T}(\mathbf{s}'_k)\, d\mathbf{s}'_k.
\end{equation}
In addition, this dimensionality-reduction transformation allows us to reformulate the original problem \eqref{op1} into a tractable form. First, we focus on the CRB matrix in the objective function \eqref{op1obj}, which depends on the desired navigation signal component and its spatial gradient at each NUE. After the channel subspace transformation in equations \eqref{subspace_w}
and \eqref{subspace_v}, $\mu_l(\mathbf{q}_l)$ and $\nabla_{\mathbf{q}_l}\mu_l(\mathbf{q}_l)$ can be re-expressed in terms of $\mathbf{b}_k$ as
\begin{equation}
\mu_{k,l}(\mathbf q_l) = \mathbf{g}_{k,l}^{N} \mathbf{b}_k,
\end{equation}
\begin{equation} \label{subspace_nabla_mu}
\nabla_{\mathbf{q}_l} \mu_{k,l}(\mathbf{q}_l) = \mathbf{D}_{k,l}(\mathbf{q}_l) \mathbf{b}_k,
\end{equation}
where
\begin{equation}
\mathbf{D}_{k,l}(\mathbf{q}_l) = \int_{\mathcal{S}'_k} \nabla_{\mathbf{q}_l} h_{k,l}^N(\mathbf{s}'_k, \mathbf{q}_l) \boldsymbol{\Phi}_k^T(\mathbf{s}'_k) d\mathbf{s}'_k.
\end{equation}
Moreover, the effective noise variance at the $l$-th NUE $\sigma_{{\rm{eff}},l}^2$ can be rewritten as
\begin{equation} \label{subspace_sigma_eff}
\sigma _{{\rm{eff}},l}^2 = \eta^\rho {\sum\limits_{m = 1}^M {\left| {\sum\limits_{k = 1}^K {{\bf{g}}_{k,l}^N{{\bf{a}}_{k,m}}} } \right|} ^2} + \sigma _{N,l}^2.
\end{equation}
Similarly, the communication achievable rate constraint (\ref{op1st1}) and the transmit power constraint (\ref{op1st2}) can be rewritten as
\begin{equation} \label{op2st1}
\frac{\left| \sum\limits_{k = 1}^K \mathbf{g}_{k,m}^{C} \mathbf{a}_{k,m}\right|^2}{\sum\limits_{i\neq m}^M \left|\sum\limits_{k = 1}^K \mathbf{g}_{k,m}^{C} \mathbf{a}_{k,i}\right|^2 + \sum\limits_{k = 1}^K\left| \mathbf{g}_{k,m}^{C} \mathbf{b}_k\right|^2 + \sigma_{C,m}^2} \geq \Gamma_m,
\end{equation}
\begin{equation} \label{op2st2}
\sum_{m=1}^M \mathbf{a}_{k,m}^H \mathbf{R}_k^P \mathbf{a}_{k,m} + \mathbf{b}_k^H \mathbf{R}_k^P \mathbf{b}_k \leq P_k^{\max},
\end{equation}
where $\Gamma_m = 2^{R_m^{\min}}-1$ and
$\mathbf{R}_k^P = \int_{\mathcal{S}'_k} \boldsymbol{\Phi}_k(\mathbf{s}'_k) \boldsymbol{\Phi}_k^H(\mathbf{s}'_k) d\mathbf{s}'_k \in \mathbb{C}^{(M+L) \times (M+L)}.$
In this context, the original infinite-dimensional problem (\ref{op1}) is reformulated as an equivalent finite-dimensional optimization problem over the coefficient vectors ${\mathbf{a}_{k,m}}$ and ${\mathbf{b}_k}$, which can be expressed as
\begin{subequations}
\label{op2}
\begin{align}
&\min_{ \{\mathbf{a}_{k,m}\}, \{\mathbf{b}_k\} } \quad \frac{1}{L} \sum_{l=1}^L \mathrm{tr}\left( \mathbf{CRB}(\mathbf{q}_l) \right) \label{op2obj} \\
&\ \ \ \ \ \ \ \mathrm{s.t.} \ \ \ \  (\ref{op2st1}), (\ref{op2st2}).\notag
\end{align}
\end{subequations}
After the channel subspace transformation, problem (\ref{op2}) is cast into a finite-dimensional form but remains non-convex due to the complicated objective function, the quadratic terms and the coupling among the optimization variables.

\subsection{Algorithm Design}
To address this non-convexity, we apply the semi-definite relaxation (SDR) technique and define matrix variables $\mathbf{A}_m = \mathbf{a}_m \mathbf{a}_m^H$ and $\mathbf{B} = \mathbf{b} \mathbf{b}^H $, where $\mathbf{a}_m = [\mathbf{a}_{1,m}^T, \dots, \mathbf{a}_{K,m}^T]^T$ and $\mathbf{b} = [\mathbf{b}_1^T, \dots, \mathbf{b}_K^T]^T$ are aggregate vectors by stacking the expansion coefficients for all $K$ satellites. Accordingly, the communication achievable rate constraint (\ref{op2st1}) is rewritten as
\begin{equation}\label{op4st1}
\mathrm{tr}(\mathbf{G}_m^C \mathbf{A}_m) \geq \Gamma_m \Bigg( \sum_{i \ne m}^M \mathrm{tr}(\mathbf{G}_m^C \mathbf{A}_i) + \mathrm{tr}(\tilde{\mathbf{G}}_m^C \mathbf{B}) + \sigma_{C,m}^2 \Bigg),
\end{equation}
where the communication channel matrices $\mathbf{G}_m^C \in \mathbb{C}^{K(M+L) \times K(M+L)}$ and $\tilde{\mathbf{G}}_{m}^{C} \in \mathbb{C}^{K(M+L) \times K(M+L)}$ for the $m$-th CUE are defined as
\begin{equation}
{ \mathbf{G}_m^C = \mathbf{g}_m^C (\mathbf{g}_m^C)^H },
\end{equation}
\begin{equation}
\tilde{\mathbf{G}}_{m}^{C} = \mathrm{blkdiag}\ \big(\mathbf{g}_{1,m}^C(\mathbf{g}_{1,m}^C)^H,\dots,\mathbf{g}_{K,m}^C(\mathbf{g}_{K,m}^C)^H\big),
\end{equation}
with $\mathbf{g}_m^C = \left[\mathbf{g}_{1,m}^{C}, \mathbf{g}_{2,m}^C, \dots, \mathbf{g}_{K,m}^C\right]^T \quad \in \mathbb{C}^{K(M+L) \times 1}$.
Similarly, the transmit power constraint (\ref{op2st2}) is updated as
\begin{equation}\label{op4st2}
\sum_{m=1}^M \mathrm{tr}(\tilde{\mathbf{R}}_k^P \mathbf{A}_m) + \mathrm{tr}(\tilde{\mathbf{R}}_k^P \mathbf{B}) \leq P_k^{\max},
\end{equation}
where $\mathbf{\tilde{R}}_k^P = \text{blkdiag}(\mathbf{0}, \dots, \mathbf{R}_k^P, \dots, \mathbf{0}) \in \mathbb{C}^{K(M+L) \times K(M+L)}$ is a block-diagonal selector matrix whose $k$-th diagonal block is the matrix $\mathbf{R}_k^P$ and all other blocks are zero, thus acting as a selector that isolates the power contribution of the $k$-th satellite only.

Furthermore, the CRB-based objective function (\ref{op2obj}) is the most challenging component to handle. To facilitate its reformulation, we define navigation channel matrix for the $l$-th NUE as $\mathbf{G}_l^N = \mathbf{g}_l^N (\mathbf{g}_l^N)^H$, where the aggregated subspace channel vector for the $l$-th NUE is given by ${\bf{g}}_l^N = [ {\bf{g}}_{1,l}^N, {\bf{g}}_{2,l}^N, \ldots, {\bf{g}}_{K,l}^N ]^T \in \mathbb{C}^{K(M+L) \times 1}$. According to equations (\ref{eq:J_def}), (\ref{eq:FIM_def}), (\ref{eq:CRB_def}), (\ref{subspace_nabla_mu}), and (\ref{subspace_sigma_eff}), the objective function (\ref{op2obj}) can be rewritten in terms of the matrix variables $\{\mathbf{A}_m\}$ and $\mathbf{B}$ as (\ref{op3}) at the top of the next page,
\begin{figure*}[!t]
\begin{equation} \label{op3}
\min_{\{\mathbf{A}_m\}, \mathbf{B}} \frac{1}{L} \sum_{l=1}^L \mathrm{tr}\left( \left( \frac{2}{\sigma_{\text{eff},l}^2} \Re\left\{ \sum_{k=1}^{K}
\mathbf D_{k,l}(\mathbf q_l)\,
\mathbf E_k \mathbf B \mathbf E_k^{H}\,
\mathbf D_{k,l}^{H}(\mathbf q_l)\right\} \right)^{-1}\right),
\end{equation}
\hrulefill
\end{figure*}
where $\sigma_{\text{eff},l}^2$ is transformed into
\begin{equation}
\sigma_{\text{eff},l}^2 = \eta^\rho \sum_{m=1}^M \mathrm{tr}(\mathbf{G}_l^N \mathbf{A}_m) + \sigma_{N,l}^2,
\end{equation}
and $\mathbf{E}_k \in \mathbb{R}^{(M+L) \times K(M+L)}$ is a block selection
matrix defined as $\mathbf{E}_k = [\mathbf{0}, \dots, \mathbf{I}_{M+L}, \dots, \mathbf{0}]$, where the identity block $\mathbf{I}_{M+L}$ is placed at the $k$-th block position, thereby selecting the coefficients corresponding to the $k$-th satellite.
To avoid explicit matrix inversion in (\ref{op3}), we introduce an auxiliary variable matrix $\mathbf{T}_l \succeq 0$ with $\mathbf{T}_l \succeq \frac{2}{\sigma_{\text{eff},l}^2} \Re\{\sum_{k=1}^{K}
\mathbf D_{k,l}(\mathbf q_l)\,
\mathbf E_k \mathbf B \mathbf E_k^{H}\,
\mathbf D_{k,l}^{H}(\mathbf q_l)\}$. According to the Schur complement theorem, the objective function (\ref{op3}) is equivalently reformulated as minimizing $\frac{1}{L} \sum_{l=1}^L \mathrm{tr}\left( \mathbf{T}_l \right)$, which is subject to the matrix inequality:
\begin{equation}\label{coupling_LMI}
\begin{bmatrix}
\mathbf{T}_l & \mathbf{I}_3 \\
\mathbf{I}_3 & \dfrac{2}{\sigma_{\text{eff},l}^2} \Re\left\{ \sum\limits_{k = 1}^K
\mathbf D_{k,l}(\mathbf q_l)\,
\mathbf E_k \mathbf B \mathbf E_k^{H}\,
\mathbf D_{k,l}^{H}(\mathbf q_l) \right\}
\end{bmatrix}
\succeq 0.
\end{equation}

Consequently, the original optimization problem (\ref{op2}) can be reformulated as
\begin{subequations}
\label{op4}
\begin{align}
&\min_{\{\mathbf{A}_m\}, \mathbf{B}, \{\mathbf{T}_l\}} \quad \frac{1}{L} \sum_{l=1}^L \mathrm{tr}\left( \mathbf{T}_l \right) \label{op4obj} \\
&\ \ \ \ \ \ \ \mathrm{s.t.} \ \ \ \  (\ref{op4st1}), (\ref{op4st2}), (\ref{coupling_LMI}),\notag\\
&\ \ \ \ \ \ \ \ \ \ \ \ \ \ \ \mathbf{A}_m \succeq 0, \quad \mathbf{B} \succeq 0, \label{op4obj3}\\
&\ \ \ \ \ \ \ \ \ \ \ \ \ \ \ \text{Rank}\left( {{{\bf{A}}_m}} \right) = 1, \quad  \text{Rank}\left( {{{\bf{B}}}} \right)= 1, \label{rank_constraint}
\end{align}
\end{subequations}
where the constraint (\ref{coupling_LMI}) is still non-convex because $\sigma_{\text{eff},l}^2$ depends on ${\mathbf{A}_m}$, which in turn induces coupling between optimize variables ${\mathbf{A}_m}$ and $\mathbf{B}$. To solve this problem, we apply a block coordinate descent (BCD) method by fixing $\sigma_{\text{eff},l}^2$ at each iteration to its value from the previous step, denoted by $\sigma_{\text{eff},l}^{2,(t-1)}$, so that \eqref{coupling_LMI} becomes a standard linear matrix inequality (LMI). In particular, a damped update strategy is employed to improve stability, i.e.,
\begin{equation}\label{update_eff}
\sigma_{\text{eff},l}^{2,(t)} = (1-\lambda')\sigma_{\text{eff},l}^{2,(t-1)} + \lambda' \sigma_{\text{eff},l}^{2,\text{new}},
\end{equation}
where $\lambda' \in (0,1)$ is a damping factor that controls the update step size and helps avoid large oscillations of the objective value across iterations, and $\sigma_{\text{eff},l}^{2,\text{new}}$ denotes the effective noise variance recomputed from the current solution ${\mathbf{A}_m^{(t)}}$.
In addition, the positive semi-definite constraint (\ref{op4obj3}) and non-convex rank-one (\ref{rank_constraint}) arise from the application of the SDR technique. To handle the non-convex rank-one constraints in (\ref{rank_constraint}), we employ a penalty-based method that incorporates these constraints into the objective function. This approach is part of a successive convex approximation (SCA) framework, where the non-convex problem is iteratively approximated by a sequence of solvable convex problems. Specifically, for a positive semi-definite matrix $\mathbf{X}$, the condition $\text{Rank}(\mathbf{X}) = 1$ is equivalent to $\mathrm{tr}(\mathbf{X}) - \lambda_{\max}(\mathbf{X}) = 0$, where $\lambda_{\max}(\cdot)$ denotes the maximum eigenvalue. Since $\mathrm{tr}(\mathbf{X}) - \lambda_{\max}(\mathbf{X})$ is always non-negative for a positive semi-definite matrix, we can penalize any deviation from the rank-one structure by adding the term $\rho \left( \sum_{m=1}^M (\mathrm{tr}(\mathbf{A}_m) - \lambda_{\max}(\mathbf{A}_m)) + (\mathrm{tr}(\mathbf{B}) - \lambda_{\max}(\mathbf{B})) \right)$ to the objective function, where $\rho > 0$ is a penalty parameter.
The resulting objective function becomes a difference-of-convex (DC) problem, since the maximum eigenvalue function $\lambda_{\max}(\cdot)$ is convex. It can be effectively handled by linearizing the concave component, i.e., $-\lambda_{\max}(\cdot)$, at each iteration. At the $t$-th iteration, we replace $\lambda_{\max}(\mathbf{A}_m)$ with its first-order Taylor expansion around the solution from the previous iteration $\mathbf{A}_m^{(t-1)}$. This is given by $\lambda_{\max}(\mathbf{A}_m^{(t-1)}) + \mathrm{tr}((\mathbf{u}_{A_m}^{(t-1)} \mathbf{u}_{A_m}^{(t-1)H}) (\mathbf{A}_m - \mathbf{A}_m^{(t-1)}))$, where $\mathbf{u}_{A_m}^{(t-1)}$ is the principal eigenvector of $\mathbf{A}_m^{(t-1)}$. Since constant terms do not affect the optimization, we can simplify the approximation of $\lambda_{\max}(\mathbf{A}_m)$ to $\mathbf{u}_{A_m}^{(t-1)H} \mathbf{A}_m \mathbf{u}_{A_m}^{(t-1)}$.
By combining the BCD method for handling the coupling in constraint (\ref{coupling_LMI}) and the SCA-based penalty method to enforce the rank-one constraint (\ref{rank_constraint}), we arrive at an iterative optimization algorithm. In the $t$-th iteration, the resulting subproblem can be written as the following convex problem:
\begin{subequations} \label{op5}
\begin{align}
    &\min_{\substack{\{\mathbf{A}_m\}, \mathbf{B}, \\ \{\mathbf{T}_l\}}} \frac{1}{L} \sum_{l=1}^L \mathrm{tr}(\mathbf{T}_l) + \rho \bigg( \sum_{m=1}^M \big( \mathrm{tr}(\mathbf{A}_m) - \mathbf{u}_{A_m}^{(t-1)H} \mathbf{A}_m \mathbf{u}_{A_m}^{(t-1)} \big) \notag \\
    & \quad \quad + \mathrm{tr}(\mathbf{B}) - \mathbf{u}_B^{(t-1)H} \mathbf{B} \mathbf{u}_B^{(t-1)} \bigg) \label{op5obj} \\
    &\ \ \ \text{s.t.} \quad (\ref{op4st1}), (\ref{op4st2}), (\ref{op4obj3}), \notag\\
    &\ \ \ \ \  \begin{bmatrix}
    \mathbf{T}_l & \mathbf{I}_3 \\
    \mathbf{I}_3 & \frac{2}{\sigma_{\text{eff},l}^{2,(t-1)}} \Re\left\{ \sum\limits_{k = 1}^K
\mathbf D_{k,l}(\mathbf q_l)\,
\mathbf E_k \mathbf B \mathbf E_k^{H}\,
\mathbf D_{k,l}^{H}(\mathbf q_l) \right\}
    \end{bmatrix} \succeq \mathbf{0}. \label{op5st1}
\end{align}
\end{subequations}
This problem is a standard semi-definite programming (SDP) formulation, which can be solved efficiently using existing convex optimization toolboxes. After obtaining the solutions $\{\mathbf{A}_m^{(t)}\}$ and $\mathbf{B}^{(t)}$, the effective noise variance $\sigma_{\text{eff},l}^{2,(t)}$ and the principal eigenvectors $\mathbf{u}_{A_m}^{(t)}$ and $\mathbf{u}_B^{(t)}$ are updated for the next iteration.
If the final solutions are not perfectly rank-one, a standard Gaussian randomization procedure can be applied to extract a high-quality rank-one approximate solution.
Once the algorithm converges to the final optimal solutions ${\mathbf{A}_m^*}$ and $\mathbf{B}^*$ from the iteration process of solving (\ref{op5}), the transmit beamformers for communication and navigation at each satellite's CAPA in the original problem (\ref{op1}) can be calculated as
\begin{equation}\label{w_final}
w_{k,m}^*(\mathbf{s}'_k) = \boldsymbol{\Phi}_k^T(\mathbf{s}'_k) \mathbf{E}_k \left( \sqrt{\lambda_{\max}(\mathbf{A}_m^*)} \, \mathbf{u}_{\mathbf{A}_m^*} \right),
\end{equation}
\begin{equation}\label{v_final}
v_k^*(\mathbf{s}'_k) = \boldsymbol{\Phi}_k^T(\mathbf{s}'_k) \mathbf{E}_k \left( \sqrt{\lambda_{\max}(\mathbf{B}^*)} \, \mathbf{u}_{\mathbf{B}^*} \right),
\end{equation}
where $\mathbf{u}_{\mathbf{A}_m^*}$ and $\mathbf{u}_{\mathbf{B}^*}$ denote the unit-norm principal
eigenvectors associated with $\lambda_{\max}(\mathbf{A}_m^*)$ and $\lambda_{\max}(\mathbf{B}^*)$, respectively.
To summarize, the overall procedure of the proposed subspace-based beamforming design for the CAPA-ICAN system in LEO satellite constellations is presented in Algorithm 1.
\begin{algorithm}
\setstretch{0.9}
\caption{Subspace-Based CAPA-ICAN Beamforming Design in LEO Satellite Constellations}
\label{alg1}
\hspace*{0.02in}{\bf Input:} $K, M, L, D_x, D_y, R_m^{\min}, P_k^{\max}, \mathbf{r}_k, \mathbf{p}_m, \mathbf{q}_l, \rho, \lambda', \varrho, \iota$.\\
\hspace*{0.02in}{\bf Output:} $w_{k,m}^*, v_k^*$.
\begin{algorithmic}[1]
\STATE{\textbf{Initialize} iteration index $t = 1$, initial feasible points $\mathbf{A}_m^{(0)}$ and $\mathbf{B}^{(0)}$.}
\REPEAT
\STATE {Update $\sigma_{\text{eff},l}^{2,(t-1)}$ according to equation (\ref{update_eff});}
\STATE {Update principal eigenvectors $\mathbf{u}_{A_m}^{(t-1)}$ and $\mathbf{u}_B^{(t-1)}$ via eigenvalue decomposition method;}
\STATE {Obtain $\mathbf{A}_m^{(t)}$ and $\mathbf{B}^{(t)}$ by solving problem (\ref{op5});}
\IF{$\mathbf{A}_m^{(t)}$ and $\mathbf{B}^{(t)}$ converge}
\IF{$\sum_{m=1}^M \left|(\mathrm{tr}(\mathbf{A}_m^{(t)}) - \lambda_{\max}(\mathbf{A}_m^{(t)}))\right| + \left| \mathrm{tr}(\mathbf{B}^{(t)}) - \lambda_{\max}(\mathbf{B}^{(t)}) \right|> \varrho$}
\STATE{Update penalty factor $\rho=\iota\rho$};
\ENDIF
\ENDIF
\STATE{Update $t=t+1$;}
\UNTIL Convergence
\STATE Obtain $w_{k,m}^*$ and $v_k^*$ by eigenvalue decomposition and matrix operations according to equations (\ref{w_final}) and (\ref{v_final}).
\end{algorithmic}
\end{algorithm}

\subsection{Algorithm Analysis}
In the following, we analyze in detail the convergence behavior and computational complexity of the proposed algorithm.

\emph{Convergence Analysis:}
For the proposed Algorithm 1, which iteratively solves the convex SDP in (\ref{op5}) to update $\{\mathbf{A}_m\}$ and $\mathbf{B}$, the objective value of the penalized problem is monotonically non-increasing. Specifically, let $\mathcal{J}^{(t)}$ denote the objective value of problem (\ref{op5}) at the $t$-th iteration. Since (\ref{op5}) is a convex SDP problem that is optimally solved at each iteration for fixed $\{\sigma_{\text{eff},l}^{2,(t-1)}\}$, $\{\mathbf{u}_{A_m}^{(t-1)}\}$, and $\mathbf{u}_B^{(t-1)}$, we have $\mathcal{J}^{(t)} \leq \mathcal{J}^{(t-1)}$ for all $t$. Moreover, the feasible set defined by the achievable rate constraint (\ref{op4st1}), transmit power constraint (\ref{op4st2}), and LMI constraint (\ref{op5st1}) is compact, which guarantees that $\{\mathcal{J}^{(t)}\}$ is bounded below. Hence, the sequence $\{\mathcal{J}^{(t)}\}$ converges by the monotone bounded criterion \cite{Convergence analyse}. In addition, according to standard results on BCD and SCA, any limit point of the generated sequence $\{\mathbf{A}_m^{(t)}, \mathbf{B}^{(t)}, \mathbf{T}_l^{(t)}\}$ satisfies the Karush-Kuhn-Tucker (KKT) conditions of the penalized SDR problem, and thus corresponds to a stationary solution of the original beamforming design. The convergence behavior of Algorithm 1 is further illustrated in Fig. \ref{Convergence}, where the average CRB is observed to decrease and stabilize within a small number of iterations for different numbers of cooperative LEO satellites.

\emph{Complexity Analysis:}
The computational complexity of Algorithm~1 is dominated by solving the convex SDP in (\ref{op5}) at each iteration. Other updates such as $\{\sigma_{\text{eff},l}^{2,(t)}\}$ and eigenvector extractions incur negligible cost. Let $N_{\mathrm{b}}=M+L$ denote the per-satellite subspace dimension and $N_{\mathrm{s}}=K N_{\mathrm{b}}$ denote the aggregate dimension. In problem (\ref{op5}), $\{\mathbf{A}_m\}$ and $\mathbf{B}$ are Hermitian positive semidefinite matrices of size $N_{\mathrm{s}}\times N_{\mathrm{s}}$, and $\{\mathbf{T}_l\}$ are Hermitian $3\times 3$ matrices. The number of decision variables is on the order of $\varsigma = \mathcal{O}\big( (M+1)N_{\mathrm{s}}^2 + 9L \big).$
Since (\ref{op5}) is a standard SDP with LMI constraints, it can be efficiently solved by an interior-point method. Problem (\ref{op5}) includes $(M+1)$ LMI constraints of dimension $N_{\mathrm{s}}$ associated with $\mathbf{A}_m\succeq\mathbf{0}$ and $\mathbf{B}\succeq\mathbf{0}$, and $L$ LMI constraints of dimension $6$ arising from (\ref{op5st1}), together with additional scalar linear constraints. Therefore, the overall barrier parameter is approximately as
\begin{equation}
\nu = \mathcal{O}\big( (M+1)N_{\mathrm{s}} + 6L \big).
\end{equation}
Following \cite{complexity analysis}, the worst-case arithmetic complexity of solving (\ref{op5}) to accuracy $\zeta>0$ in one iteration is given by
\begin{equation}
\mathcal{C}_{\mathrm{iter}} = \mathcal{O}\big( \sqrt{\nu}\,\Xi \ln(1/\zeta) \big),
\end{equation}
where $\Xi$ is a polynomial in the problem dimensions and can be upper bounded by
\begin{equation}
\Xi
= \varsigma\Big[(M+1)N_{\mathrm{s}}^3 + 216L
+ \varsigma\big((M+1)N_{\mathrm{s}}^2 + 36L\big)
+ \varsigma^2\Big].
\end{equation}
Consequently, the overall complexity of Algorithm~1 is given by
\begin{equation}
\mathcal{C}_{\mathrm{total}} = \mathcal{O}\big( I_{\mathrm{iter}} \sqrt{\nu}\,\Xi \ln(1/\zeta) \big),
\end{equation}
which grows polynomially with $K$, $M$, and $N_{\mathrm{b}}$ and is typical for SDP-based beamforming designs.

\section{Simulation Results}
This section describes the simulation setup and presents numerical results to demonstrate the effectiveness of the proposed algorithm. Without loss of generality, we consider a classical Walker Delta LEO satellite constellation similar to the Starlink system \cite{Walker Delta}.
The constellation is configured with an orbital altitude of $h^{*}=550$ km, consisting $P^{*}=72$ orbital planes, with $N^{*}=18$ satellites per plane, an inclination of $i^{*}=53^{\circ}$, and a phase factor of $F^{*}=45$. In all simulations, a subset of satellites is selected from this constellation as the service group according to an elevation-angle priority rule. The corresponding skyplot of the service group for the case with $K=5$ and an observation point located at $[6371, 0, 0]^{T}$ km in ECEF coordinate system is shown in Fig. 2. Furthermore, the CUEs and NUEs are assumed to be randomly distributed within a 30 km radius around this observation point. All integral evaluations in the simulations are performed using a 10-point Gauss-Legendre scheme \cite{Legendre}. The effective radiating area of the CAPAs is set to $A_T=0.25~\text{m}^2$ with side lengths $D_x=D_y=\sqrt{A_T}$. The noise variance is set to $\sigma_{C,m}^2=\sigma_{N,l}^2=5.6\times10^{-3}~\text{V}^2/\text{m}^2$.
Unless otherwise specified, the remaining simulation parameters are configured as summarized in Table I.

\begin{figure}[!t]
 \centering
\includegraphics [width=0.36\textwidth] {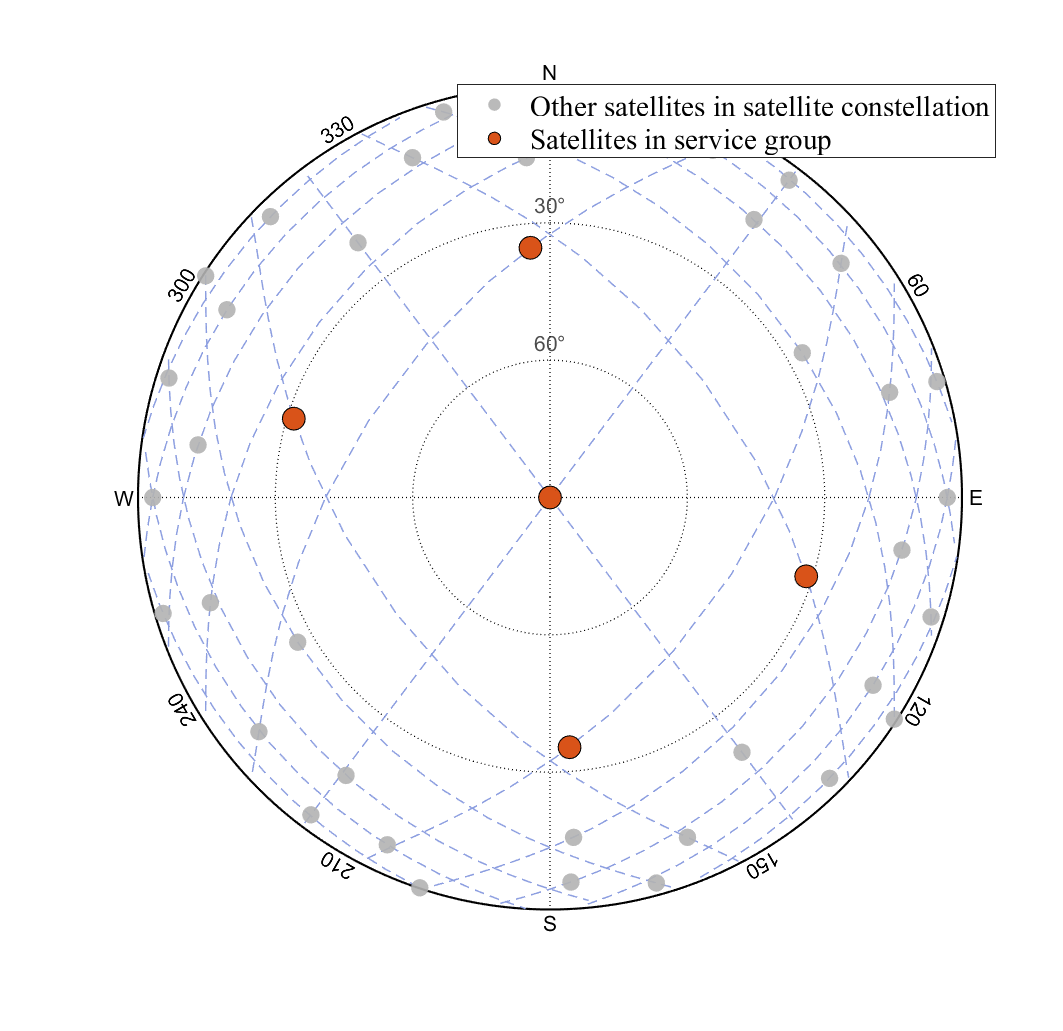}
\vspace{-7pt}
\caption {Skyplot of the service group satellites in the Walker Delta constellation.}
\label{Algorithm2}
\end{figure}

\begin{table}[!t]
\vspace{-7pt}
\small
\centering
\caption{Simulation Parameters}\label{Simulation}
\vspace{-6pt}
\begin{tabular}{|c|c|}
\hline
Parameter & Value \\ \hline
Number of service group satellites & $K=5$ \\\hline
Number of CUEs & $M=10$ \\\hline
Number of NUEs & $L=4$ \\\hline
Maximum transmit power budget & $P_k^{\max }=100~\text{mA}^2$ \\\hline
Required minimum achievable rate & $R_m^{\min}=3$ bps/Hz \\\hline
Speed of light & $c=3\times 10^{8}$ m/s \\\hline
Signal frequency & $f=35$ GHz \\\hline
Rain attenuation mean & $\mu _r=-2.6$ dB \\\hline
Rain attenuation variance & $\sigma _r^2=1.63$ dB \\\hline
Penalty factor & $\rho=1$ \\\hline
Penalty amplification factor & $\iota=1.2$ \\\hline
Damping factor & $\lambda'=0.2$ \\\hline
\end{tabular}
\end{table}

\begin{figure}[!t]
 \centering
\includegraphics [width=0.39\textwidth] {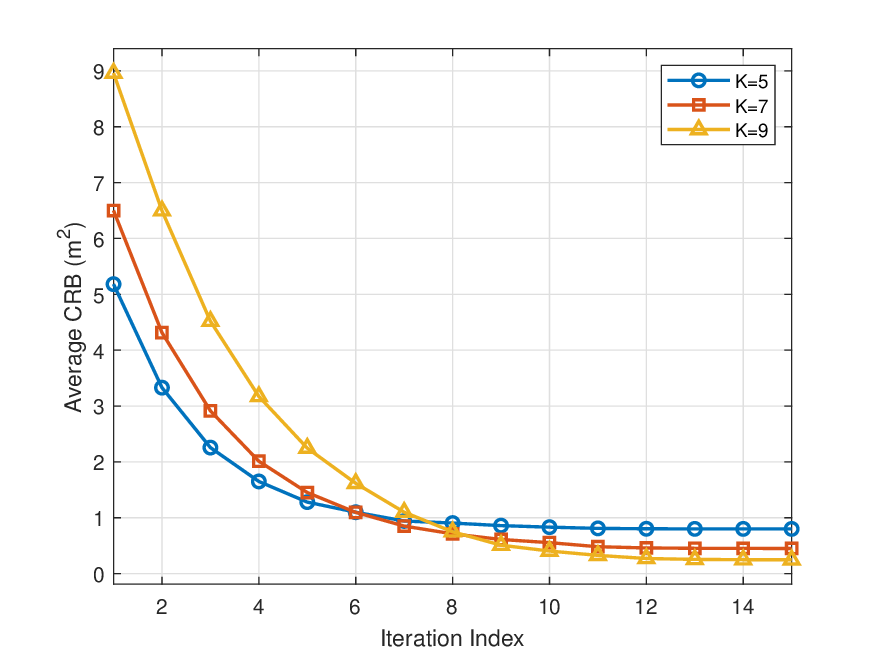}
\vspace{-7pt}
\caption {Convergence behavior of the proposed algorithm.}
\label{Convergence}
\end{figure}
First, we investigate the convergence behavior of the proposed algorithm for different numbers of cooperative satellites in the service group. In Fig. \ref{Convergence}, the vertical axis represents the objective function of the proposed algorithm, namely $\frac{1}{L} \sum_{l=1}^L \mathrm{tr}\left( \mathbf{CRB} \right)$, which is simply referred to as the average CRB. It can be observed that the objective value decreases monotonically and very rapidly during the first several iterations, and the curves become nearly flat after about 8-10 iterations, which confirms the fast convergence of the algorithm. Moreover, a larger number of cooperative satellites leads to a higher initial value because more satellites generate stronger mutual interference, resulting in a less favorable starting point. After convergence, the average CRB is significantly reduced as $K$ increases. In particular, the case with $K=9$ satellites attains the lowest average CRB, followed by the cases with $K=7$ and $K=5$, respectively, demonstrating that involving more cooperative satellites provides additional spatial diversity and design degrees of freedom, thus improving the positioning accuracy of the proposed CAPA-assisted ICAN system.

For performance comparison, we consider the following four benchmark schemes: i) Discrete Phased Array (DPA) Scheme, where each CAPA is replaced by a conventional phased array with the same physical aperture and half-wavelength element spacing, and the per-element transmit weights are optimized under the same power and rate constraints \cite{DPA}; ii) Navigation-centric Scheme, which designs the transmit beamformers by minimizing the average CRB subject only to the transmit power constraint (\ref{op1st2}) while ignoring the communication rate constraint (\ref{op1st1}); iii) Fourier-based scheme, where the continuous transmit beamformers on each CAPA are represented by Fourier series expansions and the corresponding coefficients are optimized under the same objective and constraints as the proposed algorithm \cite{Fourier-based}; iv) Zero forcing (ZF)-oriented scheme, where the transmit beamformers for communication signals designed to satisfy ZF conditions at all CUEs to cancel both inter-CUE interference and navigation interference, and the navigation component is subsequently designed under the remaining power budget \cite{ZF}.

\begin{figure}[!t]
 \centering
\includegraphics [width=0.39\textwidth] {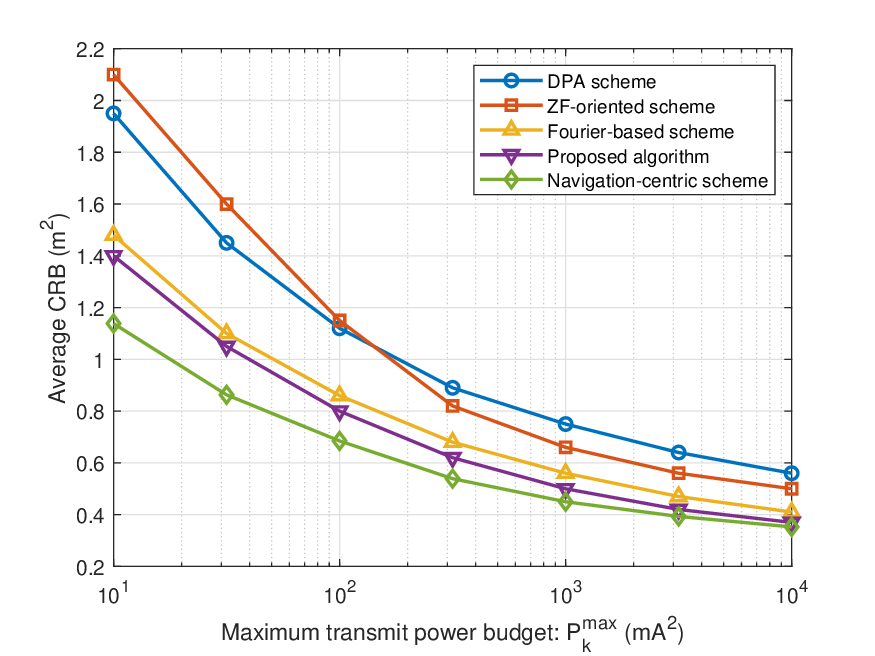}
\vspace{-7pt}
\caption {Average CRB versus the maximum transmit power budget.}
\label{compare_P}
\end{figure}
Fig. \ref{compare_P} illustrates the average CRB versus the maximum transmit power budget $P_k^{\max}$ for the proposed algorithm and the four benchmark schemes. As expected, the average CRB of all schemes decreases monotonically with increasing $P_k^{\max}$, since a larger power budget improves the received SNR at the NUEs. The Navigation-centric scheme attains the lowest CRB across the whole range, since it optimizes solely for navigation accuracy. The proposed algorithm closely follows this lower bound, especially in the medium-to-high power regime, demonstrating an effective balance between communication QoS and navigation accuracy. The Fourier-based scheme achieves a performance close to the proposed algorithm, but its reliance on high-order Fourier expansions leads to much higher computational complexity, making it less appealing in practice. In contrast, the DPA and ZF-oriented schemes exhibit noticeable performance loss. The DPA scheme is limited by the reduced spatial degrees of freedom of discrete arrays, while the ZF-oriented scheme has the highest average CRB in the low-power regime because of the stringent zero-forcing constraints. As $P_k^{\max}$ increases, the ZF-oriented scheme benefits from the larger power budget and its performance becomes closer to that of the DPA and Fourier-based schemes.

\begin{figure}[!t]
 \centering
\includegraphics [width=0.39\textwidth] {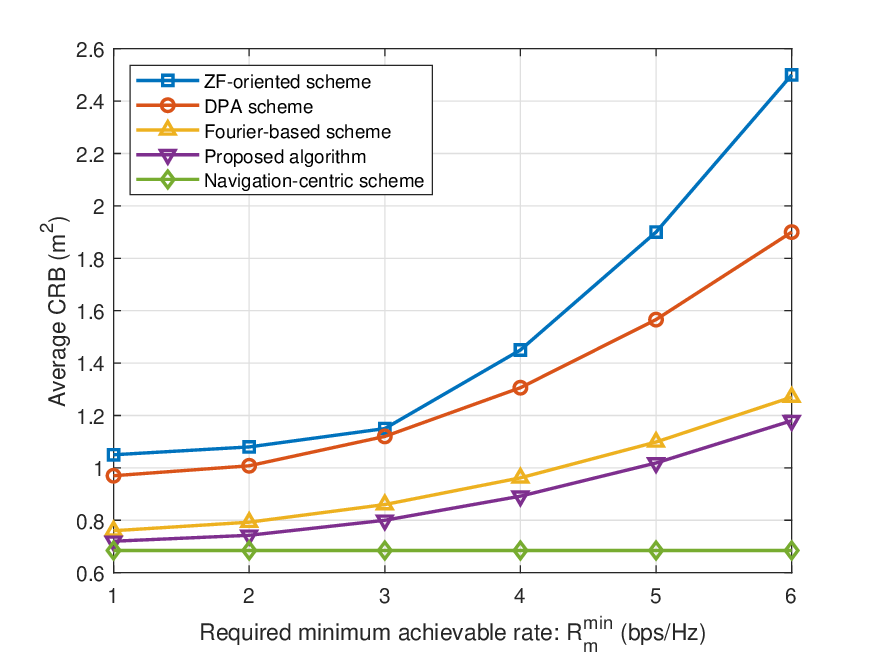}
\vspace{-7pt}
\caption {Average CRB versus the required minimum achievable rate.}
\label{compara_R}
\end{figure}
Then, Fig. \ref{compara_R} shows how the average CRB varies with the required minimum achievable rate $R_m^{\min}$. The Navigation-centric scheme is flat with respect to $R_m^{\min}$, since it ignores the rate constraint and optimizes only the navigation objective. In contrast, all rate-constrained schemes exhibit increasing CRB as $R_m^{\min}$ becomes larger. This is because a higher rate requirement forces more power and spatial degrees of freedom to be assigned to the communication links, thereby reducing the resources available for improving positioning accuracy. Among these schemes, the proposed algorithm shows the smallest growth rate, as it jointly tailors the CAPA-based signal design to accommodate both achievable rate and CRB requirements. The DPA and the ZF-oriented schemes experience the steepest increase in CRB, which is because the DPA scheme offers fewer effective spatial degrees of freedom and strict zero-forcing constraints further restrict the design space, making these two schemes particularly sensitive to high-rate requirements.

\begin{figure}[!t]
 \centering
\includegraphics [width=0.39\textwidth] {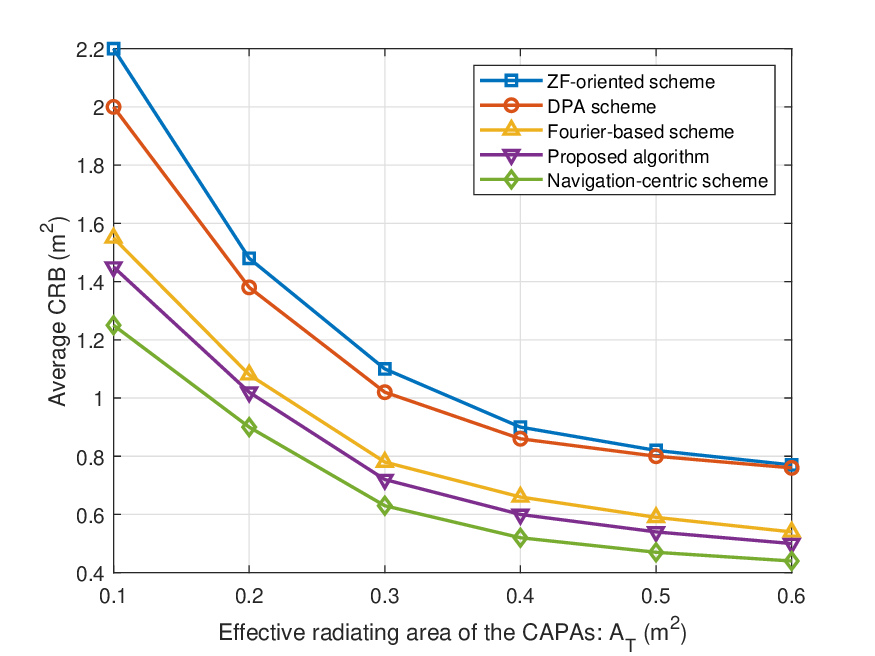}
\vspace{-7pt}
\caption {Average CRB versus the effective radiating area of the CAPAs.}
\label{compara_A}
\end{figure}
Fig. \ref{compara_A} illustrates the impact of the effective radiating area $A_T$ of the CAPAs on the average CRB. As $A_T$ increases from $0.1$ to $0.6\,\text{m}^2$, all schemes exhibit a sharp CRB reduction for small apertures and then gradually flatter, because a larger physical aperture provides higher array gain, finer angular resolution, and more orthogonal multiuser channels. The Navigation-centric scheme consistently achieves the lowest CRB, and the proposed algorithm closely tracks it. The ZF-oriented scheme yields the highest CRB across all values of $A_T$. As $A_T$ increases, its CRB decreases more rapidly and the gap to the other schemes becomes smaller. This is because a larger aperture improves channel orthogonality and makes the zero-forcing constraints easier to satisfy. On the other hand, a larger effective radiating area requires more hardware overhead and structural support, which increases payload cost and implementation complexity. Therefore, the choice of $A_T$ should balance positioning accuracy against hardware and payload cost rather than simply maximizing the aperture size.

\begin{figure}[!t]
 \centering
\includegraphics [width=0.39\textwidth] {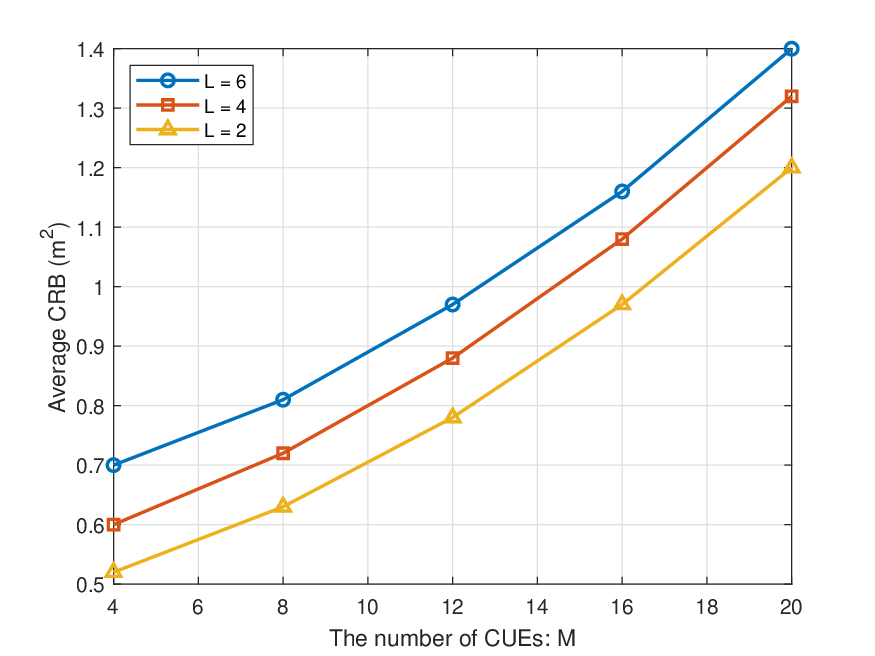}
\vspace{-7pt}
\caption {Average CRB versus the number of CUEs for different numbers of NUEs.}
\label{CUE_NUE}
\end{figure}
Next, Fig. \ref{CUE_NUE} shows the impact of the number of CUEs $M$ and NUEs $L$. It is seen that the average CRB increases monotonically and becomes steeper as $M$ grows, because serving more CUEs under the same CAPA aperture and power budget tightens the rate constraints and strengthens interference, so more power and spatial degrees of freedom must be allocated to communication rather than navigation. In addition, a larger number of NUEs leads to a higher average CRB, since the beamforming design must distribute navigation information over more spatial directions and the average is taken over a larger NUE set, including NUEs with less favorable geometries. Therefore, this figure illustrates how network density gradually degrades positioning accuracy in the proposed CAPA-assisted ICAN system.

\begin{figure}[!t]
 \centering
\includegraphics [width=0.39\textwidth] {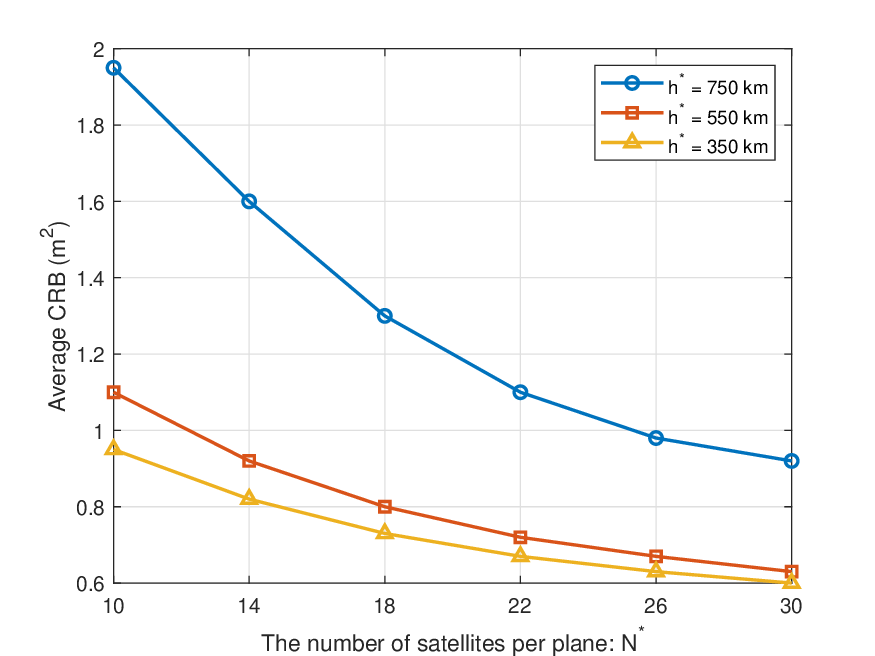}
\vspace{-7pt}
\caption {Average CRB versus the number of satellites per plane for different orbital altitudes.}
\label{satellite_fig}
\end{figure}
Lastly, Fig. \ref{satellite_fig} illustrates the impact of the orbital altitude $h^{*}$ and the number of satellites per plane $N^{*}$ on the average CRB. The average CRB drops quickly as $N^{*}$ increases and then decreases more slowly, while lower orbital altitudes consistently yield better positioning performance. This is because a denser constellation improves geometry and cooperative gain with diminishing marginal returns, and closer satellites provide stronger signals and a wider angular spread. Overall, Fig. \ref{satellite_fig} indicates that constellation design plays a crucial role and that the satellite altitude and density should be chosen by balancing deployment cost and system performance.

\section{Conclusion}
This paper proposed an ICAN framework based on LEO satellite constellations, where an electromagnetic model is established for dual-function LEO satellite equipped with CAPA. By introducing an ICAN channel subspace, the infinite-dimensional CAPA beamforming design was transformed into a finite-dimensional optimization problem that balances achievable rate and CRB. Numerical results shown that the proposed CAPA beamformer design algorithm yields notable ICAN performance gains over conventional DPA and other benchmark schemes, indicating the potential of CAPAs for future LEO satellite ICAN systems.

\begin{appendices}
\section{Derivation of $\nabla_{\mathbf{q}_l} h_{k,l}^N$ in equation (\ref{eq:grad_mu_kl})}
Herein, we provide the explicit derivation of the gradient of the navigation channel response with respect to the NUE's position, i.e., $\nabla_{\mathbf{q}_l} h_{k,l}^N(\mathbf{s}'_k, \mathbf{q}_l)$. This quantity is required both by the DPE algorithm and by the CRB evaluation in the main text.
From equation (\ref{eq:nav_channel_response}), the navigation channel response is given by $h_{k,l}^N(\mathbf{s}'_k, \mathbf{q}_l)$.
By substituting the Green's function expression in equation (\ref{eq:green_function}) into (\ref{eq:nav_channel_response}), the channel response can be factorized as
\begin{equation}
h_{k,l}^N(\mathbf{s}'_k, \mathbf{q}_l) = C_{k,l} \, f_{\text{prop}}(\mathbf{q}_l) \, f_{\text{proj}}(\mathbf{q}_l),
\end{equation}
where $C_{k,l} = -j\eta \xi_r^{1/2} e^{-j\psi_{k,l}}/(2\lambda)$ is a constant independent of $\mathbf{q}_l$. The propagation and projection terms are given by
\begin{align}
f_{\text{prop}}(\mathbf{q}_l) &= \frac{e^{-j\frac{2\pi}{\lambda} d_{k,l}}}{d_{k,l}}, \\
f_{\text{proj}}(\mathbf{q}_l) &= \mathbf{u}_l^T \left( \mathbf{I}_3 - \frac{\mathbf{v}_{k,l} \mathbf{v}_{k,l}^T}{d_{k,l}^2} \right) \mathbf{\hat{u}}_{k,y},
\end{align}
with the geometric vector $\mathbf{v}_{k,l} = \mathbf{q}_l - \mathbf{s}_k(\mathbf{s}'_k)$ and its norm $d_{k,l} = \|\mathbf{v}_{k,l}\|$. Since $\mathbf{s}_k(\mathbf{s}'_k)$ does not depend on $\mathbf{q}_l$, we have $\nabla_{\mathbf{q}_l} \mathbf{v}_{k,l} = \mathbf{I}_3$, and thus gradients with respect to $\mathbf{q}_l$ can be equivalently computed with respect to $\mathbf{v}_{k,l}$. By applying the product rule, the gradient of $h_{k,l}^N$ is given by
\begin{equation} \label{eq:app_product_rule}
\nabla_{\mathbf{q}_l} h_{k,l}^N
= C_{k,l} \big[ (\nabla_{\mathbf{q}_l} f_{\text{prop}}) f_{\text{proj}}
+ f_{\text{prop}} (\nabla_{\mathbf{q}_l} f_{\text{proj}}) \big].
\end{equation}
First, we derive the gradient of the propagation term $\nabla_{\mathbf{q}_l} f_{\text{prop}}$. The gradient of the distance $d_{k,l}$ can be expressed as
\begin{equation}
\nabla_{\mathbf{q}_l} d_{k,l} = \frac{\mathbf{v}_{k,l}}{d_{k,l}}.
\end{equation}
Using the chain rule, we obtain
\begin{equation}
\nabla_{\mathbf{q}_l} f_{\text{prop}}(\mathbf{q}_l)
= \left( -j\frac{2\pi}{\lambda d_{k,l}} - \frac{1}{d_{k,l}^2} \right)
\frac{e^{-j\frac{2\pi}{\lambda} d_{k,l}}}{d_{k,l}} \, \mathbf{v}_{k,l}.
\end{equation}
Next, we compute the gradient of the projection term $\nabla_{\mathbf{q}_l} f_{\text{proj}}(\mathbf{q}_l)$. By expanding the quadratic form, $f_{\text{proj}}(\mathbf{q}_l)$ can be rewritten as
\begin{equation}
f_{\text{proj}}(\mathbf{q}_l)
= (\mathbf{u}_l^T \mathbf{\hat{u}}_{k,y}) - \frac{N}{D},
\end{equation}
where $N = (\mathbf{u}_l^T \mathbf{v}_{k,l}) (\mathbf{v}_{k,l}^T \mathbf{\hat{u}}_{k,y})$ and $D = d_{k,l}^2 = \mathbf{v}_{k,l}^T \mathbf{v}_{k,l}$. The first term $\mathbf{u}_l^T \mathbf{\hat{u}}_{k,y}$ is constant with respect to $\mathbf{q}_l$. Since $\mathbf{v}_{k,l}$ is an affine function of $\mathbf{q}_l$ with unit Jacobian, we can equivalently compute the gradient with respect to $\mathbf{v}_{k,l}$, i.e., $\nabla_{\mathbf{q}_l} = \nabla_{\mathbf{v}_{k,l}}$. The gradients of $N$ and $D$ with respect to $\mathbf{v}_{k,l}$ are given by
\begin{align}
&\nabla_{\mathbf{v}_{k,l}} N
= \mathbf{u}_l (\mathbf{v}_{k,l}^T \mathbf{\hat{u}}_{k,y})
+ \mathbf{\hat{u}}_{k,y} (\mathbf{u}_l^T \mathbf{v}_{k,l}),\\
&\nabla_{\mathbf{v}_{k,l}} D = 2\mathbf{v}_{k,l}.
\end{align}
Applying the quotient rule to $-N/D$, we have
\begin{align}
\nabla_{\mathbf{q}_l} f_{\text{proj}}&(\mathbf{q}_l)
= - \frac{(\nabla_{\mathbf{v}_{k,l}} N) D - N (\nabla_{\mathbf{v}_{k,l}} D)}{D^2}\notag \\
&= - \frac{1}{d_{k,l}^4} \Big( d_{k,l}^2 \big[ \mathbf{u}_l(\mathbf{v}_{k,l}^T \mathbf{\hat{u}}_{k,y})
+ \mathbf{\hat{u}}_{k,y}(\mathbf{u}_l^T \mathbf{v}_{k,l}) \big]\notag  \\
& \qquad\qquad\quad - 2 (\mathbf{u}_l^T \mathbf{v}_{k,l})(\mathbf{v}_{k,l}^T \mathbf{\hat{u}}_{k,y}) \mathbf{v}_{k,l} \Big).
\end{align}
Finally, substituting the expressions of $\nabla_{\mathbf{q}_l} f_{\text{prop}}$ and $\nabla_{\mathbf{q}_l} f_{\text{proj}}$ into (\ref{eq:app_product_rule}) yields the closed-form gradient of the navigation channel response, which is expressed as equation (\ref{eq:gradient_full_expression}) at the top of the next page.
\begin{figure*}[!t]
\begin{equation} \label{eq:gradient_full_expression}
\begin{aligned}
\nabla_{\mathbf{q}_l} h_{k,l}^N (\mathbf{s}'_k, \mathbf{q}_l) = C_{k,l} \Bigg[ &\underbrace{ \left( -j\frac{2\pi}{\lambda d_{k,l}^2} - \frac{1}{d_{k,l}^3} \right) e^{-j\frac{2\pi}{\lambda} d_{k,l}} \mathbf{v}_{k,l} }_{\nabla f_{\text{prop}}} \cdot \underbrace{ \left( \mathbf{u}_l^T \left( \mathbf{I}_3 - \frac{\mathbf{v}_{k,l}\mathbf{v}_{k,l}^T}{d_{k,l}^2} \right) \mathbf{\hat{u}}_{k,y} \right) }_{f_{\text{proj}}} \\
&\quad + \underbrace{ \left( \frac{e^{-j\frac{2\pi}{\lambda} d_{k,l}}}{d_{k,l}} \right) }_{f_{\text{prop}}} \cdot \underbrace{ \left( \frac{2(\mathbf{u}_l^T\mathbf{v}_{k,l})(\mathbf{v}_{k,l}^T\mathbf{\hat{u}}_{k,y})\mathbf{v}_{k,l} - d_{k,l}^2\left[\mathbf{u}_l(\mathbf{v}_{k,l}^T\mathbf{\hat{u}}_{k,y}) + \mathbf{\hat{u}}_{k,y}(\mathbf{u}_l^T\mathbf{v}_{k,l})\right]}{d_{k,l}^4} \right) }_{\nabla f_{\text{proj}}} \Bigg].
\end{aligned}
\end{equation}
\hrulefill
\end{figure*}

\section{Proof of Theorem 1}
Let $\mathcal{C}_k$ denote the finite-dimensional ICAN channel subspace in (\ref{ICAN_channel_subspace}) and $\mathcal{C}_k^{\perp}$ be its orthogonal complement over $\mathcal{S}'_k$ under the inner product, i.e.,
\begin{equation}
\int_{\mathcal{S}'_k} f_k^*(s'_k)\,g_k(s'_k)\,\mathrm{d}s'_k = 0,\quad
f_k \in \mathcal{C}_k,\ g_k \in \mathcal{C}_k^{\perp}.
\label{eq:orth_def}
\end{equation}
Consider an arbitrary optimal solution $\{w_{k,m}(s'_k),v_k(s'_k)\}$ to problem~(32). For each $w_{k,m}(s'_k)$, take its orthogonal decomposition with respect to $\mathcal{C}_k$ and $\mathcal{C}_k^{\perp}$ as
\begin{equation}
w_{k,m}(s'_k) = w_{k,m}^{\parallel}(s'_k) + w_{k,m}^{\perp}(s'_k),
\label{eq:w_decomp}
\end{equation}
where $w_{k,m}^{\parallel}(s'_k)\in\mathcal{C}_k$ and $w_{k,m}^{\perp}(s'_k)\in\mathcal{C}_k^{\perp}$. Similarly,
\begin{equation}
v_k(s'_k) = v_k^{\parallel}(s'_k) + v_k^{\perp}(s'_k),
\label{eq:v_decomp}
\end{equation}
with $v_k^{\parallel}(s'_k)\in\mathcal{C}_k$ and $v_k^{\perp}(s'_k)\in\mathcal{C}_k^{\perp}$. For notational simplicity, the dependence on $s'_k$ will be omitted in the following, and we write $w_{k,m}$, $v_k$, $w_{k,m}^{\parallel}$, $w_{k,m}^{\perp}$, $v_k^{\parallel}$, and $v_k^{\perp}$ instead.

First, we show that the orthogonal components $w_{k,m}^{\perp}$ and $v_k^{\perp}$ do not affect any performance-relevant quantity. From the received signal expressions in (\ref{eq:comm_received_signal_decomposed}), (\ref{receive_navigation}), (\ref{eq:mu_kl_def}) and the effective noise variance in (\ref{effective_noise_variance}), every term in the communication achievable rate $R_m$, the navigation mean vectors $\boldsymbol{\mu}_l$, and the CRB matrix can be written as a linear function in the form
$\int_{\mathcal{S}'_k} h_{k,m}^C(s'_k,\mathbf{p}_m)\,w_{k,i}\,\mathrm{d}s'_k$ or $\int_{\mathcal{S}'_k} h_{k,l}^N(s'_k,\mathbf{q}_l)\,v_k\,\mathrm{d}s'_k$.
By the definition of $\mathcal{C}_k$, the conjugate kernels $h_{k,m}^C(\cdot,\mathbf{p}_m)^*$ and $h_{k,l}^N(\cdot,\mathbf{q}_l)^*$ belong to $\mathcal{C}_k$. Together with equation \eqref{eq:orth_def}, we have
\begin{align}
\int_{\mathcal{S}'_k} h_{k,m}^C(s'_k,\mathbf{p}_m)\,w_{k,i}^{\perp}\,\mathrm{d}s'_k &= 0,
\label{eq:orth_comm}\\
\int_{\mathcal{S}'_k} h_{k,l}^N(s'_k,\mathbf{q}_l)\,v_k^{\perp}\,\mathrm{d}s'_k &= 0.
\label{eq:orth_nav}
\end{align}
Hence, all linear functions depend only on $w_{k,m}^{\parallel}$ and $v_k^{\parallel}$. As a result, replacing each $w_{k,m}$ by $w_{k,m}^{\parallel}$ and each $v_k$ by $v_k^{\parallel}$ leaves $R_m$, $\boldsymbol{\mu}_l$, and $\mathbf{CRB}$ unchanged.

Next, we examine the transmit power. According to equation \eqref{eq:power_derivation_b} and the orthogonal decompositions, we have
\begin{align}
\int_{\mathcal{S}'_k} |w_{k,m}|^2\,\mathrm{d}s'_k
&= \int_{\mathcal{S}'_k} \big(|w_{k,m}^{\parallel}|^2 + |w_{k,m}^{\perp}|^2\big)\,\mathrm{d}s'_k,
\label{eq:norm_w}\\
\int_{\mathcal{S}'_k} |v_k|^2\,\mathrm{d}s'_k
&= \int_{\mathcal{S}'_k} \big(|v_k^{\parallel}|^2 + |v_k^{\perp}|^2\big)\,\mathrm{d}s'_k.
\label{eq:norm_v}
\end{align}
Therefore, discarding $w_{k,m}^{\perp}$ and $v_k^{\perp}$ does not increase the transmit power, while preserving feasibility and the objective value of \eqref{op1}.

Finally, let $\tilde w_{k,m}=w_{k,m}^{\parallel}$ and $\tilde v_k=v_k^{\parallel}$. Then $\{\tilde w_{k,m},\tilde v_k\}$ is feasible for \eqref{op1}, attains the same achievable rates and CRB as the original optimal solution, and satisfies $\tilde w_{k,m}(s'_k),\tilde v_k(s'_k)\in\mathcal{C}_k$. Therefore, an optimal solution with all beamformers lying in $\{\mathcal{C}_k\}$ always exists, which proves Theorem~1.

\end{appendices}

\end{document}